\documentclass[a4paper]{article}
\usepackage{a4wide}

\usepackage{amssymb}
\usepackage{latexsym}

\newcommand{\R}{\mathbb{R}}
\usepackage{natbib}
\setcitestyle{numbers,square}

\usepackage{amsmath}
\usepackage{mathtools}

\usepackage[utf8]{inputenc}

\usepackage{url}
\usepackage{svg}
\usepackage{todonotes}

\usepackage{csquotes}

\usepackage{xspace}
\makeatletter
\xspaceaddexceptions{\grqq \grq \csq@qclose@i \} }
\makeatother

\newcommand{\eg}{e.\,g.,\ }
\newcommand{\ie}{i.\,e.,\ }
\newcommand{\cf}{cf.\ }

\newcommand{\discotec}{DisCoTec\xspace}
\newcommand{\gene}{GENE\xspace}
\newcommand{\selalib}{SeLaLib\xspace}

\usepackage[capitalise,nameinlink,noabbrev]{cleveref}

\usepackage[hypcap=true]{subcaption}

\IfFileExists{pgfplots.sty}{
    \usepackage{pgfplots}
    \usepackage{pgfplotstable}
    \pgfplotsset{compat=newest}
}{}

\usepgfplotslibrary{units} 
\usepgfplotslibrary{groupplots} 
\pgfplotsset{ 
    errorBars/.style={
            error bars/error bar style={
                    thick,
                },
            error bars/x dir=none,
            error bars/y dir=both,
            error bars/y explicit,
        },
    timeplot/.style={
        legend style={cells={align=center}},
        label style={align=center},
        width=\linewidth, 
        grid=major, 
        grid style={dashed,gray!30}, 
        xlabel=Simulation time $t$, 
        legend pos = north east,
        table/x index={0},
        table/col sep=space,
        filter discard warning=false,
        no markers,
        unbounded coords = jump,
        table/col sep=space,
        xmin=0
    },
}
\usepgfplotslibrary{colorbrewer} 
\usetikzlibrary{colorbrewer} 
\pgfplotsset{cycle list/Dark2}
\tikzset{%
    fg/.style={solid, mark=square,mark options=solid,color=Dark2-A},
    hat/.style={densely dotted, semithick, mark=triangle,mark options=solid,color=Dark2-B},
    fullweighting/.style={densely dashdotted, mark=x,mark options=solid, color=Dark2-C},
    biorthogonal/.style={dashed, mark=pentagon,mark options=solid, color=Dark2-D},
    twodim/.style={cyan,mark color=cyan},
    threedim/.style={blue,mark color=blue},
    fourdim/.style={green,mark color=green},
    fivedim/.style={purple, mark color=purple},
    sixdim/.style={orange, mark color=orange},
}
\pgfplotsset{
    discard if not/.style 2 args={
        x filter/.append code={
            \edef\tempa{\thisrow{#1}}
            \edef\tempb{#2}
            \ifx\tempa\tempb
            \else
                
            \fi
        }
    }
}
\IfFileExists{pgfplotstable.sty}{
    \usepackage{pgfplotstable}
}{}
\usepackage{xintexpr}

\usepackage{float}
\usepackage{tikz}
\usetikzlibrary{backgrounds,arrows,shapes,fit,shadows,decorations.markings,matrix,positioning,decorations.pathreplacing,calligraphy}%
\pgfdeclarelayer{background}    
\pgfsetlayers{background,main}  

\usepgfplotslibrary{external}
\makeatletter
\renewcommand{\todo}[2][]{\tikzexternaldisable\@todo[#1]{#2}\tikzexternalenable}
\makeatother

\pgfplotsset{
    PosStyle/.style={
            xmin=-0.1, xmax=1.1, samples=101,
            cycle list name=color list,
            legend cell align=left,
            thick, domain=0:1,
            font=\small, align=left,
            legend style={cells={align=left}}
        },
    SymStyle/.style={
            PosStyle, xmin=-3.2, xmax=3.2, domain=-3.2:3.2, no markers, yticklabel style={right},
        },
    BiorthogonalStyle/.style={
	       	PosStyle, xmin=-1.2, xmax=2.2, domain=-1.2:2.2, no markers, yticklabel style={right}, xtick={-1,0,...,2}, ytick={-0.25,0,...,0.75},
        },
    WeightingStyle/.style={
    	PosStyle, xmin=-1.2, xmax=2.2, domain=-1.2:2.2, no markers, yticklabel style={right}, xtick={-1,0,...,2}, ytick={-5,0,...,5},
    },
}

\tikzstyle{smalllegend}=[font=\tiny, align=left]

\tikzset{declare function={
            hat(\i,\h,\x)= max(1-abs((\x / \h) - \i),0);
        }}

\tikzset{declare function={
            preprewavelet(\i,\h,\x)=-0.5*hat(-1+\i,\h,x)+hat(0+\i,\h,x)-0.5*hat(1+\i,\h,x);
        }}

\tikzset{declare function={
            biorthogonal_2_2(\i,\h,\x)=-0.125*hat(-2+\i,\h,x)-0.25*hat(-1+\i,\h,x)
            +0.75*hat(0+\i,\h,x)
            -0.25*hat(1+\i,\h,x)-0.125*hat(2+\i,\h,x);
        }}

\newcommand{\Z}{\mathbb{Z}}
\newcommand{\NN}{\mathbb{N}}
\newcommand{\proj}{\mathcal{P}}
\newcommand{\LR}{{L^2(\R)}}
\newcommand{\LRnorm}[1]{{\norm{#1}_{L^2}}}
\newcommand{\scaprod}[2]{\langle #1, #2 \rangle}
\newcommand{\diffd}{\text{d}}
\newcommand{\vect}[1]{{\boldsymbol{#1}}}
\newcommand{\setI}{\mathcal{I}}
\newcommand{\snorm}[1]{{\left\lvert #1 \right\rvert }}

\newcommand{\scalarl}{\ensuremath{l}} 
\newcommand{\justl}{\ensuremath{\vect{\scalarl}}} 
\newcommand{\lmin}{\ensuremath{\justl^\text{min}}} 
\newcommand{\lmax}{\ensuremath{\justl^\text{max}}} 

\newcommand{\uana}{\ensuremath{u_\text{exact}}\xspace} 
\newcommand{\uct}{\ensuremath{u_\text{ct}}} 
\newcommand{\ufgmax}{\ensuremath{u_\text{fg,\lmax}}} 

\newcommand{\vecx}{\ensuremath{\vect{x}}\xspace} 
\newcommand{\dt}{\ensuremath{\Delta t}\xspace} 

\usepackage[mode=text,group-four-digits=true,group-separator={,},binary-units=true]{siunitx}
\catcode`\%=12\relax
\DeclareSIUnit[number-unit-product = ]\percent{
\catcode`\%=14\relax

\DeclareSIUnit{\dof}{DOF}

\usepackage[thinc]{esdiff}
\providecommand{\abs}[1]{\left\lvert#1\right\rvert}                    
\providecommand{\norm}[1]{\ensuremath{\left\lVert#1\right\rVert}}                    
\providecommand{\Lone}[1]{\ensuremath{\norm{#1}_1}}                                 

\providecommand{\deriv}[3][]{
  \ensuremath{\frac{\partial^{#1} {#2}}{\partial {#3}^{#1}}}}
\providecommand{\grad}{\ensuremath{\nabla}}

\newcommand{\hbasis}{\ensuremath{\psi}}
\newcommand{\hbasishat}{\ensuremath{\hbasis^\mathtt{hat}}\xspace}
\newcommand{\hbasisbiorthogonal}{\ensuremath{\hbasis^\mathtt{bo}}\xspace}
\newcommand{\hbasisfullweighting}{\ensuremath{\hbasis^\mathtt{fw}}\xspace}

\usetikzlibrary{positioning, calc, decorations.pathreplacing}
\usetikzlibrary{pgfplots.colormaps, math}
\pgfplotsset{compat=1.15}

\usepackage{xifthen}

\usepackage{amsfonts} 
\DeclareMathSymbol{\shortminus}{\mathbin}{AMSa}{"39}

\newcommand{\symbolplus}[2]{
    \ifthenelse{\equal{#2}{0}}
			{#1}
			{
				\ifthenelse{\cnttest{#2}<0}
							{{#1}{#2}}
							{{#1}+{#2}}
			}
}

\tikzset{stencilcoefficient/.style={draw,solid,fill=white,inner sep=0.5,align=center,minimum size=1.4em},
    stencilhighlight/.style={rectangle,rounded corners,fill=red!15,fill opacity=0.5,thick,inner sep=3pt}}

\begin{document}

\title{A mass-conserving sparse grid combination technique with biorthogonal hierarchical basis functions for kinetic simulations}%
\author{Theresa {Pollinger}, Johannes {Rentrop}, Dirk {Pfl\"uger}, Katharina {Kormann}}
\maketitle

\begin{abstract}
The exact numerical simulation of plasma turbulence is one of the assets and challenges in fusion research.
For grid-based solvers, sufficiently fine resolutions are often unattainable due to the curse of dimensionality.
The sparse grid combination technique provides the means to alleviate the curse of dimensionality for kinetic simulations.
However, the hierarchical representation for the combination step with the state-of-the-art hat functions suffers from poor conservation properties and numerical instability.

The present work introduces two new variants of hierarchical multiscale basis functions for use with the combination technique: the biorthogonal and full weighting bases.
The new basis functions conserve the total mass and are shown to significantly increase accuracy for a finite-volume solution of constant advection.
Further numerical experiments based on the combination technique applied to a semi-Lagrangian Vlasov--Poisson solver show a stabilizing effect of the new bases on the simulations. 
\end{abstract}

\section{The curse of dimensionality in kinetic simulations}
The simulation of the time evolution of plasma in magnetic confinement fusion devices is an important part of plasma fusion research, and as such, serves both as an asset and a challenge.
While a magnetohydrodynamic (MHD) description of the plasma can capture many properties, important phenomena---as for instance the Landau damping phenomenon---are not present in the MHD model.
In this case, the more comprehensive kinetic description in phase-space is necessary.
Such a description is computationally much more demanding due to the phase-space being six dimensional.
Under certain conditions, the gyrokinetic and the drift-kinetic models---that reduce the dimensionality to five or four---provide an accurate description with a reduced dimensionality in velocity space. 
In addition to the high dimensionality, a further challenge stems from the fact that important plasma features can occur on small scales, in a way that they only start to appear at fine resolutions in the simulation.
The so-called curse of dimensionality---the exponential increase in the number of degrees of freedom (\si{\dof}) when increasing the dimensionality---haunts any approach that fits a regular grid into the phase space for direct simulation.
Taking these two properties (high dimensionalities and small scales) together, it becomes obvious that simulations that are resolved finely enough to have predictive quality can be too large in terms of compute time and/or memory footprint even for today's largest compute systems.

One approach to alleviate the curse of dimensionality is to use the sparse grid combination technique. It combines differently resolved numerical solutions multiple times throughout the course of the simulation. 
The combination technique allows to re-use existing solvers in a black-box fashion and adds an extra level of parallelism to the parallelism that may already be present in the solver~\cite{lago_exahd_2020}.
For instance, previous work applying the gyrokinetic solver \gene~\cite{jenko_global_2013} has shown that the eigenvalues in the linear part of the simulation can be obtained by a combination technique solution~\cite{Kowitz13Sparse}.
However, current approaches usually did not include the conservation of invariants---such as mass---and even had to deal with a more general problem:
When simulating up to the nonlinear quasi-stationary turbulent phase of an ITG instability scenario, the \gene combination technique might become numerically unstable~\cite[p. 319]{lago_exahd_2020}.
The effect was more prevalent when using shorter recombination intervals, a greater number of grids, or higher resolutions~\cite[section 3.6.4]{obersteiner_spatially_2021}.
This presented a severe problem for running large-scale combination technique plasma instability simulations at physically meaningful resolutions.

This work now employs two solvers that are easier to analyze numerically than \gene because they preserve known invariants: a finite volume solver and a semi-Lagrangian Vlasov solver implemented in the \selalib library~\cite{kormann_massively_2019} that both conserve mass up to machine precision.
With the finite volume solver we consider a simple advection equation where the solution is known analytically, which provides a verifiable baseline.
For the semi-Lagrangian solver we then study two benchmark problems from plasma physics: the Landau damping phenomenon and the two-stream plasma instability.

We introduce two variants of mass-conserving hierarchical multiscale bases that turn the combination step into a mass-conserving operation and establish the connection to sparse grid and combination technique theory.
Both types of basis function are rooted in the theory of biorthogonal~\cite{cohen_biorthogonal_1992} and lifting~\cite{sweldens_lifting_1998} wavelets.
The exact hierarchization and dehierarchization steps are illustrated, and we draw parallels to related work that solves the Vlasov equation directly in the hierarchical sparse grid space~\cite{haefele_parallel_2005,besse_wavelet-mra-based_2008,deriaz_six-dimensional_2018,deriaz_divergence-free_2006,kormann_sparse_2016,guo_adaptive_2017,tao2019sparse} or using the combination technique~\cite{kowitz_combination_2013,lago_exahd_2020}.
The mass-conserving basis functions are shown to reduce the error between combination technique and full grid solution in the advection scenario.
When using an equal number of degrees of freedom, the mass-conserving combination technique delivers results that are comparable in accuracy to a full grid solution in the cases of Landau damping and two-stream instability. 
Furthermore, the most important finding is that the mass-conserving bases provide stability in combination technique scenarios where the current state of the art, based on interpolets, becomes numerically unstable.
This fact may permit future exa-scale plasma simulations to draw from all benefits of the sparse grid combination technique.

\subsection{Related work}
There is a wide field of closely and loosely related work touching on the main topic of this paper: using sparse-grid techniques to efficiently compute solutions for kinetic problems.
A necessary distinction is whether the numerical scheme (differential operators, integral operators, time-stepping) is formulated in terms of the multiscale basis functions, or if the combination technique---also known as Smolyak sparse grids---is used, which allows to have the solver operate with an arbitrary basis representation.

The works by \citeauthor{haefele_parallel_2005}~\cite{haefele_parallel_2005} and \citeauthor{besse_wavelet-mra-based_2008}~\cite{besse_wavelet-mra-based_2008} fall into the first category.
These works are particularly related to our work, since they use a semi-Lagrangian scheme for the solution of the Vlasov--Poisson system, employing interpolet---a.k.a.~hat---basis functions.
A similar approach is taken by \citeauthor{deriaz_six-dimensional_2018}~\cite{deriaz_six-dimensional_2018}, using interpolets with finite difference schemes.
Earlier work by \citeauthor{deriaz_divergence-free_2006}~\cite{deriaz_divergence-free_2006} focuses on divergence- and curl-free---\ie conservative---schemes with wavelet constructions for flow simulations.
While neither of the mentioned lines of work explicitly uses a sparse grid construction, they employ adaptive hierarchical schemes, which will result in basis constructions that are similar to spatially adaptive hierarchical sparse grid bases~\cite{Pflueger10Spatially}.
\citeauthor{kormann_sparse_2016}~\cite{kormann_sparse_2016} implemented hierarchical basis operators for the Vlasov--Poisson equations into the \selalib~\cite{kormann_massively_2019} code, using the classical hat functions and higher order polynomials.
The work introduces a multiplicative $\delta$f ansatz to cope with the problem of approximating Gaussians with the sparse grid technique.
More recently, a line of work by \citeauthor{guo_adaptive_2017}~\cite{guo_adaptive_2017,tao2019sparse} successfully used an adaptive multiwavelet basis for Discontinuous Galerkin simulations of the Vlasov--Poisson and Vlasov--Maxwell systems.
Notably, the multiwavelet basis also conserves momenta up to the chosen order of the DG basis.
\citeauthor{griebel_multiscale_2003}~\cite{griebel_multiscale_2003,koster_multiskalen-basierte_2002} employed spatially adaptive sparse grids with biorthogonal basis constructions---like $\hbasisbiorthogonal$ introduced in \cref{subsubsec:biorthogonalbases}---for the simulation of turbulent flows.

The second category, combination technique based solvers, employs the hierarchical basis only between the solver steps to exchange the information between the component grids.
For flow simulations, \citeauthor{huber_turbulenzsimulation_1996}~\cite{huber_turbulenzsimulation_1996} formulated custom hierarchization and dehierarchization operators to make the simulation stable.
These operators were problem and domain specific, taking into account the physical properties to be conserved and the geometry of the domain in the simulation.

Previous work for plasma simulations could extrapolate eigenvalues in the linear part of the simulation by way of the combination technique~\cite{Kowitz13Sparse}.
The combination technique---with some adaptation---also performed well for linearized gyrokinetic simulations~\cite{kowitz_combination_2013}.
For turbulent plasma simulations however, as discussed before, the combination technique applied to \gene would often become numerically unstable~\cite[p. 319]{lago_exahd_2020}.

Our new results are based on the semi-Lagrangian approach to the Vlasov--Poisson system; in contrast to \eg \citeauthor{haefele_parallel_2005}~\cite{haefele_parallel_2005} and \citeauthor{deriaz_six-dimensional_2018}~\cite{deriaz_six-dimensional_2018}, this work solves the systems of equations not directly in the hierarchical basis, but employs the combination technique to couple solvers for structured grids.
It introduces biorthogonal linear wavelets as intermediate representations in the combination technique, since the interpolet basis is not mass-conserving and can lead to the simulation becoming unstable.
The results presented here may allow to revisit some of the aforementioned experiments with relatively little overhead at improved stability and accuracy.

To the best of our knowledge, so far no attempts have been made at higher-dimensional combination technique time-dependent simulations with bases other than hierarchical hat functions.

\section{Wavelets, hierarchical basis and basis transformations}

To introduce the necessary concepts used in this paper, we give a short overview on the theory of multiresolution analysis and the wavelet transform~\cite{cohen2003numerical}.
In this context, we will introduce biorthogonal wavelet bases~\cite{cohen_biorthogonal_1992}, and, as a special case, the hierarchical hat function basis. We consider an unbounded domain in this section and discuss boundary conditions for compact domains in section~\ref{sec:compact_domain}.

\subsection{Multiresolution analysis and biorthogonal wavelets}
A multiresolution analysis on $\LR$ is a hierarchy of function spaces
\begin{equation}
\dots V_{-2} \subset V_{-1} \subset V_{0} \subset V_{1} \subset V_{2} \subset \dots \subset \LR
\end{equation} with the following properties:
\begin{itemize}
	\item The union of all function spaces is dense in $\LR$ and there is no redundancy:
	\begin{equation}
	\overline{\cup_{l \in \Z} V_l } = \LR \ , \quad \cap_{l \in \Z} V_l = \{ 0 \} \ .
	\end{equation}
	\item Each space is spanned by dilates/translates of a \textit{scaling function} $\phi$ (it is possible to have a set of scaling functions, but we focus on the case with only one scaling function):
	\begin{equation}
	\phi_{ls}(x) := \phi (2^l x - s) \ , \quad V_l = \text{span} \{ \phi_{ls}\}_{s \in \Z} \ .
	\end{equation}
\end{itemize}
One can think of the series of spaces as successive approximation spaces for a function $f \in \LR$. Each projection $\proj_{l} f \in V_l$ is an approximation with resolution $2^{-l}$.
We call $l$ the \textit{level} of the approximation. 
The level $l=0$ corresponds to the coarsest possible level on the unit interval, which will be discussed later.

Since $\phi \in V_0 \subset V_1 = \text{span} \{ \phi_{1s}\}_{s \in \Z}$ the scaling function has to satisfy the multiscale equation
\begin{equation}
\phi(x) = \sum_{s \in \Z} h_s \phi(2x - s)  \label{eq:scalingrelation}
\end{equation}
with coefficients $h_s \in \R$, which explains its name.

Let us consider the complement space of $V_{l-1}$ in $V_{l}$, also known as the \textit{hierarchical increment space} $W_l$
\begin{equation}
V_l = V_{l-1} \oplus W_l \ , \qquad V_{l-1} \cap W_l = \{0\} \ . \label{eq:spacedecomp}
\end{equation}
Applying this relation iteratively leads to the decomposition (setting $W_0 \coloneqq V_0$)
\begin{equation}
V_l = \oplus_{l'=0}^l W_{l'} \ , \qquad \LR = \overline{\oplus_{l'=0}^\infty W_{l'}} \ .
\end{equation}

We aim to define basis functions $\psi_{ls}$ for $W_l$, so that $W_l = \text{span}\{ \psi_{ls}\}_{s \in \Z}$. These functions will be called \textit{wavelets}. Then, every function $f \in \LR$ can be represented as
\begin{equation}
f = \sum_{l=0}^\infty \sum_{s \in \Z} \alpha_{ls} \psi_{ls} \ ,
\end{equation}
where we set $\psi_{0s} \coloneqq \phi_{0s}$. We call $\alpha_{ls}$ the \textit{hierarchical} or \textit{wavelet coefficients}.

However, the space $W_l$ is not uniquely defined, it needs further specification.
Requiring it to be the orthogonal complement, for example, leads to \textit{orthonormal wavelets} \cite{daubechies1988}.
Here, instead, we only demand that the $\psi_{ls}$ are dilates/translates of a \textit{mother wavelet} $\psi$, i.e. $\psi_{ls}(x) \coloneqq \psi(2^{l-1} x - s)$, in particular $\psi_{10} = \psi$.
Since $W_1 \subset V_1$, there are---similar to \cref{eq:scalingrelation}---coefficients $g_s \in \R$, so that
\begin{equation}
\psi(x) = \sum_{s \in \Z} g_s \phi(2x - s) \ . \label{eq:scalingpsi}
\end{equation}

In order to uniquely define the mother wavelet $\hbasis$, we employ \textit{biorthogonal wavelets}.
For this purpose, we require a \textit{dual} multiresolution analysis with a dual scaling function $\tilde{\phi}$, a dual mother wavelet $\tilde{\psi}$ and coefficients $\tilde{h}_s, \tilde{g}_s$, so that
\begin{equation}
\tilde{\phi}(x) = 2 \sum_{s \in \Z} \tilde{h}_s \tilde{\phi}(2x - s) \ , \qquad \tilde{\psi}(x) = \sum_{s \in \Z} \tilde{g}_s \tilde{\phi}(2x - s) \ . \label{eq:scalingtilde}
\end{equation}

Theorem 3.8 in \cite{cohen_biorthogonal_1992} states that, if $(h_s)$ and $(\tilde{h}_s)$ are finite sequences with $\sum_{s} h_s \tilde{h}_{s-2k} = \delta_{k0}$, if the Fourier transforms of the corresponding $\phi$ and $\tilde{\phi}$ are bounded and in $\LR$,
and if we define
\begin{equation}
g_s \coloneqq (-1)^{1-s} \tilde{h}_{1-s} \ , \qquad \tilde{g}_s \coloneqq (-1)^{1-s} h_{1-s} \ , \label{eq:defgs}
\end{equation}
then the $\psi_{ls}$, $\tilde{\psi}_{ls}$ constitute two dual frames, so that for any $f \in \LR$,
\begin{equation}
f = \sum_{l=0}^\infty \sum_{s \in \Z} 2^{l} \scaprod{\tilde{\psi}_{ls}}{f} \psi_{ls} = \sum_{l=0}^\infty \sum_{s \in \Z} 2^{l} \scaprod{\psi_{ls}}{f} \tilde{\psi}_{ls} \ .
\end{equation}
Furthermore, if and only if $\int \phi(x) \tilde{\phi}(x-s) \diffd x = \delta_{s0}$, the $\psi_{ls}$, $\tilde{\psi}_{ls}$ are two dual Riesz bases and we have the following biorthogonality relations
\begin{gather}
	\begin{aligned}
		\scaprod{\phi_{ls}}{\tilde{\phi}_{ls'}} &= 2^{-l} \delta_{ss'} \ , &  \scaprod{\psi_{ls}}{\tilde{\psi}_{l's'}} &= 2^{-l} \delta_{ll'} \delta_{ss'} \ ,  \\
		\scaprod{\phi_{(l-1)s}}{\tilde{\psi}_{ls'}}&= 0 \ , & \scaprod{\psi_{ls}}{\tilde{\phi}_{(l-1)s'}} &= 0 \ ,
	\end{aligned} \label{eq:biorthogonal}
\end{gather}
where $\scaprod{\cdot}{\cdot}$ is the $L^2$ scalar product.
Note that we use different normalization factors compared to the original paper.

In practice, we are interested in calculating the \textit{wavelet transform} or \textit{hierarchization} of a function whose approximation in $V_l$ is given by $f_l \coloneqq \proj_l f = \sum_{s \in \Z} c_{ls} \phi_{ls}$. This amounts to computing $\alpha_{l's} = 2^{l'}\scaprod{\tilde{\psi}_{l's}}{f_l}$ for $0 \leq l' \leq l$, $s \in \Z$.
One starts by choosing the two sequences $(h_s)$ and $(\tilde{h}_s)$ in such a way that they meet the above mentioned conditions. The scaling functions and wavelets do not have to be known explicitly.
From an algorithmic point of view, it suffices to know $(h_s)$, $(\tilde{h}_s)$ and, by consequence, $(g_s)$, $(\tilde{g}_s)$ in order to perform the well-known wavelet transform:

Due to \cref{eq:spacedecomp} we can expand $f_l$ as
\begin{equation}
f_l = \sum_{s \in \Z} c_{ls} \phi_{ls} = \sum_{s \in \Z} c_{(l-1)s} \phi_{(l-1)s} + \sum_{s \in \Z} \alpha_{ls} \psi_{ls} \ . \label{eq:hybridrepsingle}
\end{equation}
Taking the scalar product with $2^l \tilde{\psi}_{lk}$ and using \cref{eq:scalingtilde} and \cref{eq:biorthogonal} we obtain
\begin{equation}
\begin{aligned}
\alpha_{lk} &= 2^{l} \scaprod{\tilde{\psi}_{lk}}{f_l} = 2^{l} \sum_{s \in \Z} c_{ls} \scaprod{\tilde{\psi}_{lk}}{\phi_{ls}} =  \sum_{s \in \Z} \sum_{s' \in \Z} c_{ls} \tilde{g}_{s'} 2^l \scaprod{\tilde{\phi}_{l(s'+2k)}}{\phi_{ls}} = \sum_{s \in \Z} \sum_{s' \in \Z} c_{ls} \tilde{g}_{s'} \delta_{(s'+2k)s}\\
&= \sum_{s \in \Z} c_{ls} \tilde{g}_{(s-2k)} \ .
\end{aligned}
\label{eq:filter1}
\end{equation}

Analogously, taking the scalar product with $2^{l-1}\tilde{\phi}_{(l-1)k}$, we get
\begin{equation}
\begin{aligned}
c_{(l-1)k} &= 2^{l-1} \scaprod{\tilde{\phi}_{(l-1)k}}{f_l} = 2^{l-1} \sum_{s \in \Z} c_{ls} \scaprod{\tilde{\phi}_{(l-1)k}}{\phi_{ls}} = \sum_{s \in \Z} \sum_{s' \in \Z} c_{ls} \tilde{h}_{s'} 2^l \scaprod{\tilde{\phi}_{l(s'+2k)}}{\phi_{ls}} \\
& = \sum_{s \in \Z} c_{ls} \tilde{h}_{(s-2k)} \ .
\end{aligned}
\label{eq:filter2}
\end{equation}
Replacing $l$ by $l-1$, these two steps can be applied iteratively down to $l'=0$.
Alternatively, the transform could just be carried through to some other level $l^\text{min} \geq 0$ in order to retain a hybrid representation of $f_l$.
This is a beneficial property when discussing the sparse grid combination technique.

Equations \eqref{eq:filter1} and \eqref{eq:filter2} have the structure of a convolution with the \textit{filter mask} $(\tilde{g}_s)$ resp. $(\tilde{h}_s)$, which is displaced with a stride of two. $(\tilde{h}_s)$ covers the even indices, while $(\tilde{g}_s)$ covers the odd ones (due to the shift of one in Definition \eqref{eq:defgs}).

The inverse wavelet transform or \textit{dehierarchization} is obtained by taking the scalar product of \eqref{eq:hybridrepsingle} with $2^l \tilde{\phi}_{lk}$,
\begin{equation}
\begin{aligned}
c_{lk} &=  2^l \scaprod{\tilde{\phi}_{lk}}{f_l} = \sum_{s \in \Z} c_{(l-1)s} 2^l \scaprod{\tilde{\phi}_{lk}}{\phi_{(l-1)s}} + \sum_{s \in \Z} \alpha_{ls}2^l \scaprod{\tilde{\phi}_{lk}}{\psi_{ls}} \\
&= \sum_{s \in \Z} c_{(l-1)s} h_{k-2s} + \sum_{s \in \Z} \alpha_{ls}g_{k-2s} \ ,
\end{aligned}
\label{eq:invfilter}
\end{equation}
which, again, can be computed iteratively from level zero or some $l^\text{min}$ up to $l$.
Due to \cref{eq:filter1,eq:filter2,eq:invfilter}, $(\tilde{h}_s)$ and $(\tilde{g}_s)$ are called \textit{decomposition filters} while $(h_s)$ and $(g_s)$ are called \textit{reconstruction filters}.

\subsection{Relevant bases for this paper}
The following three types of hierarchical basis functions will be used throughout this paper.
They all employ the classical hat function, either as scaling function or dual scaling function,
\begin{equation}
\phi(x) = \max(1 - \abs{x}, 0) \ .
\end{equation}
This function has compact support, which leads to finite filter lengths, and it is interpolating, thus $\phi(s) = \delta_{s0}$ for $s \in \Z$.
The latter implies that the coefficients for the representation of $f_l \in V_l$ in terms of the $\phi_{ls}$ are just the function values at the grid points, i.e. $c_{ls} = f(2^{-l}s)$.
Since the hat function is piecewise linear, $f_l$ will be a linear spline approximating $f$ in $V_l$. It is easy to verify from the multiscale equation \eqref{eq:scalingrelation} that the corresponding filter coefficients are given by
\begin{equation}
(h_s)_{-1 \leq s \leq 1} = (\tfrac{1}{2}, 1, \tfrac{1}{2}) \ .
\end{equation}

\begin{figure}[h!]
  \centering
	\begin{tikzpicture}[]
	\begin{axis}[SymStyle]
	\addplot[samples at={-4,-1, 0, 1, 4}] {hat(0, 1, x)};
	\addplot[hat, samples at={-4, 0, 0.5, 1, 4}] {max(1 - abs(2*x - 1), 0)};
	\addlegendentry[smalllegend]{$\phi(x) = \max(1 - \abs{x}, 0)$}
	\addlegendentry[smalllegend]{$\hbasishat = \phi(2x - 1)$}
	\end{axis}
	\end{tikzpicture}
	\caption{\enquote{Hat} function: scaling function for piecewise linear multi-resolution analysis, and standard choice for hierarchical basis $\hbasishat$.}
	\label{fig:sg:hat}
\end{figure}
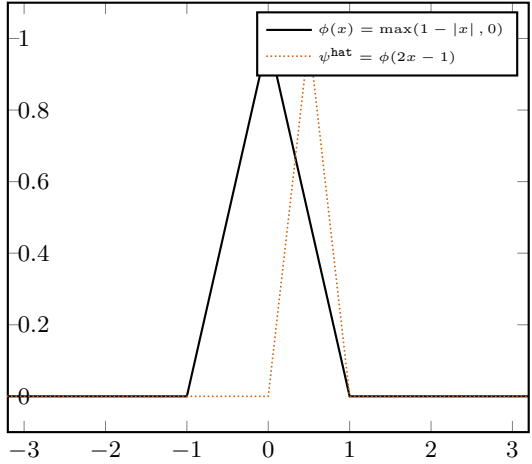

\subsubsection{Hierarchical hat basis}
The hierarchical hat basis is the standard basis used in most of the sparse grid literature.
This is why we use it in this paper as a point of reference.
Outside of the sparse grid literature, it is often framed as the interpolet basis of first order~\cite{deslauriers_symmetric_1989}.
Here, the dual scaling function is the Dirac distribution $\tilde{\phi}(x) = \delta(x)$~\cite{sweldens_lifting_1998}. In this case $\tilde{\phi} \notin \LR$ and all scalar products have to be understood as a dual pairing. Also, this violates the requirements of Theorem 3.8 in~\cite{cohen_biorthogonal_1992} and one can show that the hierarchical hat basis is indeed not a Riesz basis.

The Dirac distribution satisfies the functional equation $\delta(x) = 2 \delta(2x)$, which immediately serves as its multiscale equation. Hence, the corresponding filters have only one non-vanishing entry, $\tilde{h}_s = \delta_{s0}$. It follows from \cref{eq:defgs} that $g_{s} = \delta_{s1}$. Therefore, the wavelets themselves are also hat functions that live at odd grid points, and we will refer to them as $\hbasishat$ (cf. \cref{fig:sg:hat}).
In addition, the second step of the decomposition, \cref{eq:filter2}, becomes redundant. The $c_{ls}$ stay the same throughout the levels.
This makes the hierarchical basis transform especially efficient (\enquote{lazy wavelets}, \cf\cite{sweldens_lifting_1998}). In summary, the full set of filters is given by
\begin{equation}
\begin{aligned}
(h_s)_{-1 \leq s \leq 1}^\mathtt{hat} &= (\tfrac{1}{2}, 1, \tfrac{1}{2}) \ , & (g_s)_{0 \leq s \leq 2}^\mathtt{hat} &= (0, 1, 0) \ , \\
(\tilde{h}_s)_{-1 \leq s \leq 1}^\mathtt{hat} &= (0, 1, 0) \ , & (\tilde{g}_s)_{0 \leq s \leq 2}^\mathtt{hat} &= (-\tfrac{1}{2}, 1, -\tfrac{1}{2}) \ .
\end{aligned}
\end{equation}

\begin{figure}[h!]
  \centering
	\begin{subfigure}[]{0.45\textwidth}
    \centering
		\begin{tikzpicture}[]
		\begin{axis}[BiorthogonalStyle]
		\addplot[biorthogonal, no markers] {biorthogonal_2_2(1, 0.5, x)};
		\addlegendentry[smalllegend]{$\hbasisbiorthogonal$}
		\end{axis}
		\end{tikzpicture}
  \end{subfigure}
	\begin{subfigure}[]{0.45\textwidth}
    \centering
		\begin{tikzpicture}[]
		\begin{axis}[WeightingStyle]
		\addplot[fullweighting, no markers] table {data/dualwavelet.dat};
		\addlegendentry[smalllegend]{$\hbasisfullweighting$}
		\end{axis}
		\end{tikzpicture}
	\end{subfigure}
  \caption{The two choices of mass-conserving wavelet functions: $\hbasisbiorthogonal$ to the left and $\hbasisfullweighting$ to the right.}
  \label{fig:sg:biorthogonalbases}
\end{figure}

\subsubsection{2,2-biorthogonal bases}\label{subsubsec:biorthogonalbases}
The simplest biorthogonal wavelet basis that uses the hat function as scaling function, is the example with $(N,\tilde{N})=(2,2)$ in chapter 6.A of \cite{cohen_biorthogonal_1992}. We will denote it by $\hbasisbiorthogonal$.
The full set of filters is given by
\begin{equation}
\begin{aligned}
(h_s)_{-1 \leq s \leq 1}^\mathtt{bo} &= (\tfrac{1}{2}, 1, \tfrac{1}{2}) \ , & (g_s)_{-1 \leq s \leq 3}^\mathtt{bo} &= (-\tfrac{1}{8}, -\tfrac{1}{4}, \tfrac{3}{4}, -\tfrac{1}{4}, -\tfrac{1}{8}) \ , \\
(\tilde{h}_s)_{-2 \leq s \leq 2}^\mathtt{bo} &= (-\tfrac{1}{8}, \tfrac{1}{4}, \tfrac{3}{4}, \tfrac{1}{4}, -\tfrac{1}{8}) \ , & (\tilde{g}_s)_{0 \leq s \leq 2}^\mathtt{bo} &= (-\tfrac{1}{2}, 1, -\tfrac{1}{2}) \ .
\end{aligned}
\label{eq:filtersbo}
\end{equation}

The third type of basis we use in this paper is almost identical to the last example.
Only the roles of $\phi$ and $\tilde{\phi}$ are interchanged.
This means that the decomposition filters become the analysis filters and vice versa:
\begin{equation}
\begin{aligned}
(h_s)_{-2 \leq s \leq 2}^\mathtt{fw} &= (-\tfrac{1}{4}, \tfrac{1}{2}, \tfrac{3}{2}, \tfrac{1}{2}, -\tfrac{1}{4}) \ , & (g_s)_{0 \leq s \leq 2}^\mathtt{fw} &= (-\tfrac{1}{2}, 1, -\tfrac{1}{2}) \ , \\
(\tilde{h}_s)_{-1 \leq s \leq 1}^\mathtt{fw} &= (\tfrac{1}{4}, \tfrac{1}{2}, \tfrac{1}{4}) \ , & (\tilde{g}_s)_{-1 \leq s \leq 3}^\mathtt{fw} &= (-\tfrac{1}{8}, -\tfrac{1}{4}, \tfrac{3}{4}, -\tfrac{1}{4}, -\tfrac{1}{8}) \ .
\end{aligned}
\label{eq:filtersfw}
\end{equation}
We call this version the \textit{full weighting} basis, $\hbasisfullweighting$, because the decomposition filter $(\tilde{h}_s)$ in this case is known from the full weighting restriction in the multigrid literature. 

\begin{figure}[h!]
	\centering
	\begin{subfigure}[]{\textwidth}
		\begin{subfigure}{0.45\textwidth}
			\includegraphics[width=\textwidth]{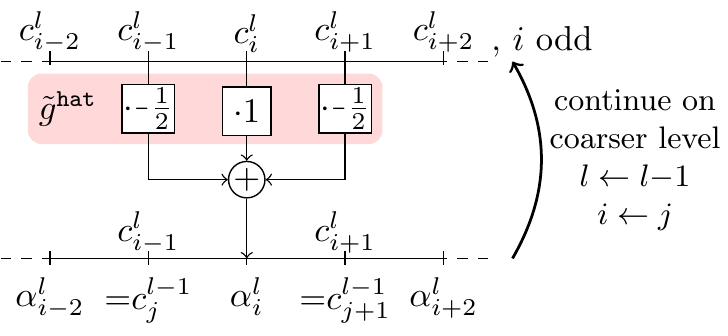}
			\caption{Hierarchical hat basis: For the hierarchization, new values on the coarser level are identical to the previous ones.}
			\label{fig:conserving:stencil:hat}
		\end{subfigure}\hfill
		\begin{subfigure}{0.45\textwidth}
			\includegraphics[width=\textwidth]{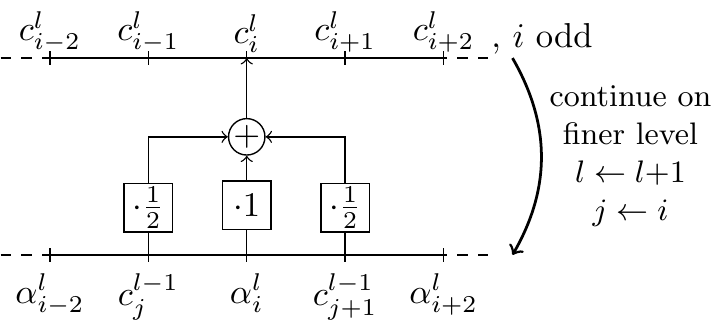}
			\caption{Hierarchical hat basis: Coarser values are kept, finer values are obtained by the sum of interpolation and hierarchical surplus.}
			\label{fig:conserving:stencild:hat}
		\end{subfigure}
	\end{subfigure}\\
	\begin{subfigure}[]{\textwidth}
		\begin{subfigure}[]{0.45\textwidth}
			\includegraphics[width=\textwidth]{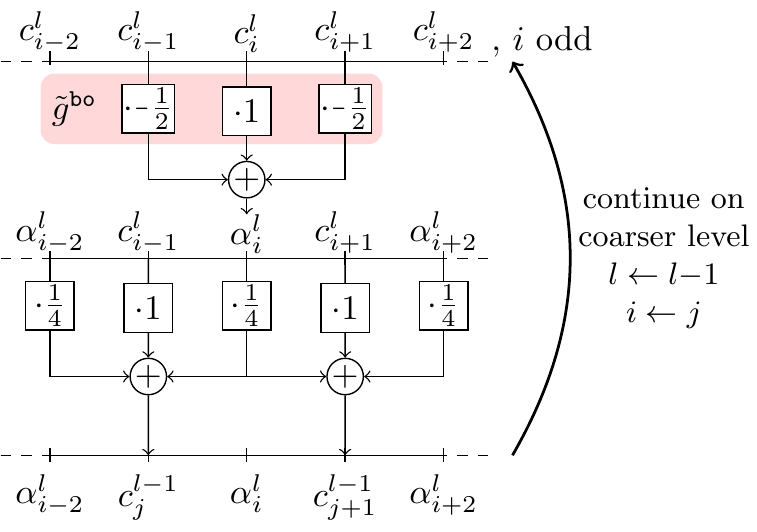}
			\caption{Biorthogonal basis: Hierarchical coefficients are calculated as before, but then the scaling function coefficients are updated, too.}
			\label{fig:conserving:stencil:biorthogonal}
		\end{subfigure}\hfill
		\begin{subfigure}[]{0.45\textwidth}
			\includegraphics[width=\textwidth]{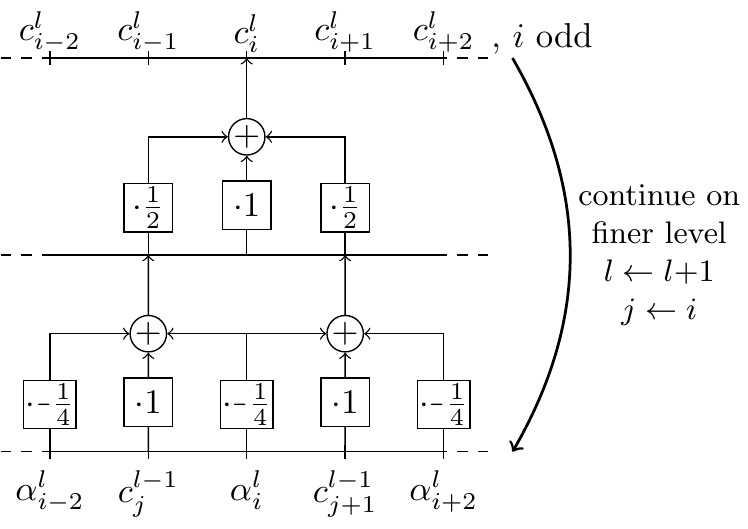}
			\caption{Biorthogonal basis: The pattern of inverse signs for odd-numbered stencil positions compared to the hierarchization is visible.}
			\label{fig:conserving:stencild:biorthogonal}
		\end{subfigure}
	\end{subfigure}\\
	\begin{subfigure}[]{\textwidth}
		\begin{subfigure}[]{0.45\textwidth}
			\includegraphics[width=\textwidth]{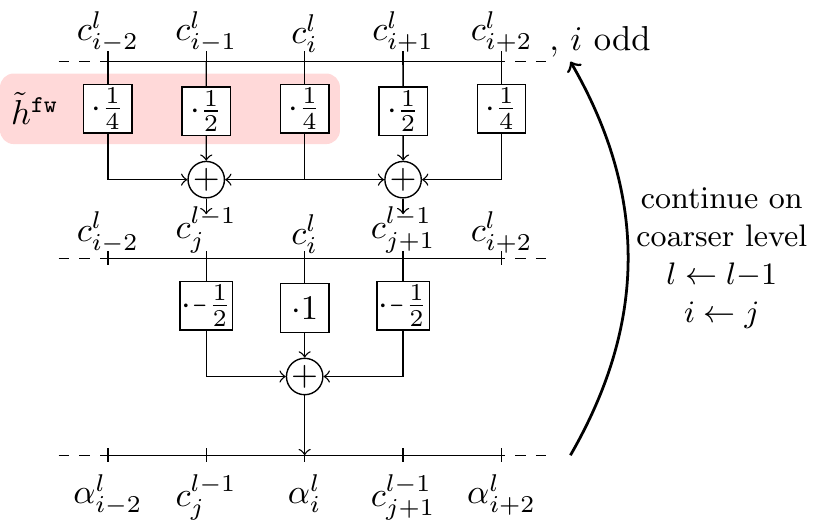}
			\caption{Full weighting basis: The filters are virtually the same as in the biorthogonal case, but applied in reverse order.}
			\label{fig:conserving:stencil:fullweighting}
		\end{subfigure}\hfill
		\begin{subfigure}[]{0.45\textwidth}
			\includegraphics[width=\textwidth]{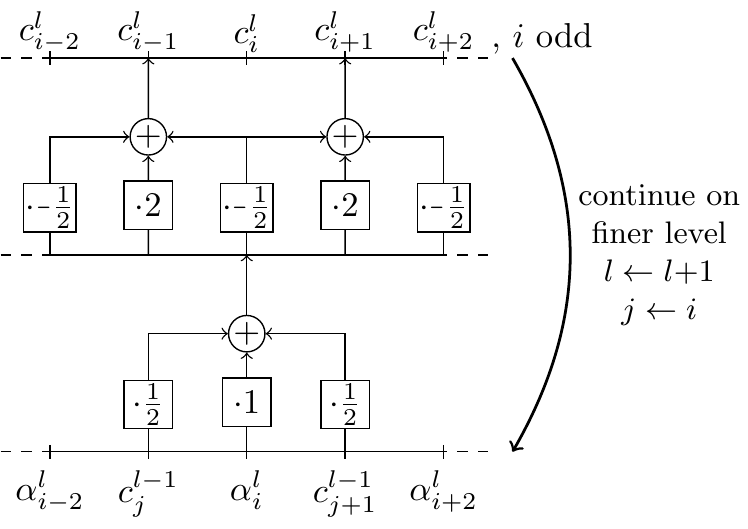}
			\caption{Full weighting basis: One can observe the \enquote{lifting} property, allowing to greedily compute the transform in-place.}
			\label{fig:conserving:stencild:fullweighting}
		\end{subfigure}
	\end{subfigure}
	\caption{Filters for the hierarchization and dehierarchization operations for the different multiscale bases. The hierarchization (left column) is described by \cref{eq:filter1,eq:filter2}; the dehierarchization (right column) is described by \cref{eq:invfilter}. In Figures (c) to (f), filters from \cref{eq:filtersbo,eq:filtersfw} that are not explicitly depicted arise from the subsequent execution of the two lifting steps.}
	\label{fig:conserving:stencils}
\end{figure}

The shapes of $\hbasisbiorthogonal$ and $\hbasisfullweighting$ can be seen in \cref{fig:sg:biorthogonalbases}. $\hbasisbiorthogonal$ is a linear B-spline wavelet just like $\hbasishat$, while $\hbasisfullweighting$ is the wavelet that serves as the dual for $\hbasisbiorthogonal$.
Although the filters in the biorthogonal case are wider than for $\hbasishat$, they can be computed in-place with two data sweeps per level instead of one, where the intermediate filters are always three elements wide.
This is due to their possible decomposition as lifting wavelets, as discussed in \cite{sweldens_lifting_1998}.
\Cref{fig:conserving:stencils} displays the hierarchization and dehierarchization operations resulting from the bases introduced in this section, and highlights the decomposition and reconstruction filters where applicable.

\subsection{Conservation of mass}
For our application, the motivating advantage of the biorthogonal bases over the hierarchical basis is the fact that they have a vanishing integral or zeroth moment.
For the two cases presented in this work, this can easily be shown by \cref{eq:scalingpsi}, together with the fact that in both cases $\phi$ is integrable (as proved in \cite{cohen_biorthogonal_1992}) and $\sum_{s \in \Z} g_s = 0$.
This implies for the hybrid representation
\begin{equation}
f_l = \sum_{s \in \Z} c_{l^\text{min}s} \phi_{l^\text{min}s} + \sum_{l^\text{min}<l' \leq l} \sum_{s \in \Z} \alpha_{l's} \psi_{l's} \label{eq:hybridrep}
\end{equation}
that the zeroth moment is fully concentrated in the minimal level and, consequently, that it remains the same throughout all approximations, independent of the level.

In a physical sense, this signifies the conservation of mass throughout the levels of approximation.
Since the conservation of mass is a key property of the type of PDEs considered in this paper, this can potentially play an important role when applying the sparse grid combination technique, \cf \cref{subsec:sg:combinationtechnique}.
The reason is that the latter is based on an extrapolation using approximations on different levels.
Conservation of mass throughout the levels results in a conserved mass for the extrapolant.

Conversely, this is not the case when using the hierarchical hat basis.
In this scenario, the wavelets themselves are hat functions with a non-vanishing integral, so the contributions to the integral are spread among the levels.
As an extreme example, consider a function whose mass is distributed very granularly, so that all $c_{l^\text{min}s}$ and $\alpha_{ls}$ for $l < L$ are zero.
All approximations with $l < L$ will carry no mass and the extrapolant will effectively consist of only one partial approximation at level $L$.

This leads to the hypothesis that the biorthogonal wavelet bases are better suited for the context of mass conservation than the hierarchical basis.
\Cref{sec:experiments} is going to validate this proposition by three numerical studies.

\subsection{Wavelets on compact domains}\label{sec:compact_domain}
In this work, only functions on compact domains will be considered, which can be scaled to the unit interval $[0,1]$ or the hypercube $[0,1]^d$ in $d$ dimensions.
Usually, one would need to make boundary adjustments for the basis functions close to the boundary in order to account for the desired regularity properties, when faced with reduced stencil sizes due to missing neighboring points outside the domain.
For instance, a suitable boundary treatment for $\hbasisbiorthogonal$ is derived in \cite{koster_multiskalen-basierte_2002}.
In all cases considered here, we will use periodic boundary conditions, which means that the boundaries can be treated like the boundaries at internal partitions of the solver.
This removes the need for boundary adjusted basis functions.

\section{Sparse grids and the combination technique}
In this chapter we introduce sparse grids and, based on that, the sparse grid combination technique \cite{Bungartz04Sparse}.
The concept is based on the hierarchical subspace splitting presented in the last chapter.

\subsection{Sparse grids}
Sparse grids can be seen as a compression technique for the representation of multidimensional functions.
Consider a function $f:[0,1]^d \to \R$.
In order to discretize it, we choose a hybrid wavelet representation similar to \cref{eq:hybridrep}.
We use tensor products of the one-dimensional approximation spaces to generalize them to multiple dimensions.
For example, the two-dimensional version of the subspace decomposition in \eqref{eq:spacedecomp} reads as
\begin{equation}
\begin{aligned}
V_{l_1,l_2} &= V_{l_1} \otimes V_{l_2} = (V_{l_1 - 1} \oplus W_{l_1}) \otimes (V_{l_2 - 1} \oplus W_{l_2}) \\
&= (V_{l_1 - 1}\otimes V_{l_2 - 1}) \oplus (V_{l_1 - 1}\otimes W_{l_2}) \oplus (V_{l_2 - 1} \otimes W_{l_1}) \oplus (W_{l_1} \otimes W_{l_2}) \ .
\end{aligned}
\end{equation}
Again, this can iteratively be expanded into hierarchical increment spaces down to a minimal level.
In order to account for the remaining cross terms between $V$ and $W$, we define the multidimensional scaling resp. wavelet functions as
\begin{equation}
\phi_{\vect{l} \vect{s}} (\vect{x}) = \prod_{i=1}^{d} \phi_{l_i s_i} (x_i) \ , \qquad \psi_{\vect{l} \vect{s}} (\vect{x}) = \prod_{i=1}^{d} \psi'_{l_i s_i} (x_i) \ , \qquad \psi'_{l_i s_i} = \begin{cases} \phi_{l_i s_i} & l_i = l_i^\text{min} \\ \psi_{l_i s_i} & l_i > l_i^\text{min} \end{cases} \ ,
\end{equation}
where bold indices are multi-indices, e.g.~$\vect{l} = (l_1,\dotsc,l_d)$.
The multidimensional (de-)hierarchization operation reduces to a sequential execution of one-dimensional (de-)hierarchization operations.

The classical discretization on a \textit{full} grid, i.e. a regular rectangular grid of level $\vect{l}$, is then given by
\begin{equation}
f_\vect{l} = \sum_\vect{s} c_{\vect{l}^\text{min} \vect{s}} \phi_{\vect{l}^\text{min} \vect{s}} + \sum_{ \vect{l}^\text{min} \leq \vect{l'} \leq \vect{l} } \sum_\vect{s} \alpha_{\vect{l'} \vect{s}} \psi_{\vect{l'} \vect{s}} \ . \label{eq:fullgridfunct}
\end{equation}
Binary operators between multi-indices are understood component-wise.
Also, whenever the first term is written out for clarity, it should implicitly be excluded from the sum in the second term.
The number of basis functions and thus the number of grid points in each direction is given by $N_i \coloneqq 2^{l_i} + 1$ (including both boundaries).
For an isotropic discretization ($N_i = N$) the total number of grid points is then given by $N^d$.
The exponential dependency on the number of dimensions $d$ is often referred to as the curse of dimensionality.

In order to lift this curse, we introduce the so called \textit{generalized sparse grid} approximation
\begin{equation}
f^{(\text{s})}_{\vect{l}} = \sum_\vect{s} c_{\vect{l}^\text{min} \vect{s}} \phi_{\vect{l}^\text{min} \vect{s}} + \sum_{\vect{l'} \in \setI_\vect{l}} \sum_\vect{s} \alpha_{\vect{l'} \vect{s}} \psi_{\vect{l'} \vect{s}} \ . \label{eq:sparsefunct}
\end{equation}
Here, the second sum does not run over all levels up to $\vect{l}$ but, instead, the levels are chosen from an index set $\setI_\vect{l}$.
This index set is usually designed to exclude many of the levels with high isotropic resolution compared to the full grid, while most of the anisotropic levels are kept.
This reduces the number of grid points drastically.
The resulting \textit{target level} $\vect{l}$ of the set is defined by the maximal occurring $l'_i$ in each direction.

The specific design of the index set depends on the decay behavior of the wavelet coefficients, which is linked to the regularity properties of the considered function class.
It is also possible to let the index set be chosen by an adaptive algorithm~\cite{gerstner2003dimension,Pflueger10Spatially}.
The only requirement for the index set is to be downward closed, i.e. $\forall \vect{l'} \in \setI_\vect{l}: \vect{l}^\text{min} \leq \vect{j} \leq \vect{l'} \Rightarrow \vect{j} \in \setI_\vect{l}$, which ensures that the (de-)hierarchization operations exist.

Let us, as an example, consider the classical sparse grid for an isotropic target resolution $\vect{l} = (n,\dotsc,n)$ for some $n \in \NN$, and denote it by $f^{(\text{s})}_{n}$.
The corresponding level set is given by
\begin{equation}
	\setI_n = \left\{ \vect{l'} \in \NN^d : \snorm{\vect{l'}}_1 \leq n \right\}
	\label{eq:classicalsg}
\end{equation}
with $\vect{l}^\text{min} = (0,\dotsc,0)$ and the $\ell^1$ norm $\snorm{\cdot}_1$.
Geometrically, this level set is merely a simplex in $\NN^d$ compared to the full hypercube that corresponds to \cref{eq:fullgridfunct}.
This choice is optimal for the approximation in the $L^2$ norm of functions from Sobolev spaces with dominating mixed smoothness, $f \in H^2_\text{mix}$.
This Sobolev space consists of all functions whose mixed weak derivatives up to second order are bounded in $L^2$,
\begin{equation}
	H^2_\text{mix} \coloneqq \left\{ f \in L^2 : \norm{f}_{H^2_\text{mix}} < \infty \right\} \ , \qquad \norm{f}_{H^2_\text{mix}} \coloneqq \left( \sum_{\snorm{\vect{m}}_\infty \leq 2 } \norm{D^{\vect{m}} f}^2_{L^2} \right)^{1/2} \ ,
\end{equation}
where $D^{\vect{m}} f \coloneqq \tfrac{\partial^{\snorm{\vect{m}}_1} }{\partial^{m_1}_{x_1} \cdots \partial^{m_d}_{x_d}}$ is the $\vect{m}$-th weak derivative.

For the case of the hierarchical basis, it can be shown \cite{Bungartz04Sparse} that the bound for the representation error is given by ($N\coloneqq 2^n$)
\begin{equation}
	\LRnorm{f - f^{(\text{s})}_{n}} \lesssim N^{-2} (\log N)^{d-1} \norm{f}_{H^2_\text{mix}} \ ,
\label{eq:reperror}
\end{equation}
which is only worse by a logarithmic factor compared to the usual $\mathcal{O}(N^{-2})$ for a piece-wise linear approximation on a full grid.
In contrast, the number of points on the sparse grid behaves asymptotically like $\mathcal{O}(N (\log N)^{d-1})$.
This is orders of magnitude less than in the full grid case, since the exponential dependency on the dimension is only present in the $\log$-Term.

\subsection{The combination technique}\label{subsec:sg:combinationtechnique}
The combination technique (CT) is an alternative representation of a sparse grid function \eqref{eq:sparsefunct}.
The goal is to represent the sum over hierarchical increment spaces by a linear combination of full grid functions \eqref{eq:fullgridfunct}, which can be decomposed into the necessary hierarchical subspaces, i.e., all spaces in the index set $\setI$.
Since many subspaces on lower levels will then be counted multiple times, one has to account for it by subtracting suitable full grid functions containing those subspaces.
This is similar to the inclusion exclusion principle from combinatorics.
For example, for the classical sparse grid \eqref{eq:classicalsg} in 2D this leads to the combination formula
\begin{equation}
f^{(\text{s})}_{n} = \sum_{l_1 + l_2 = n} f_{l_1,l_2} - \sum_{l_1 + l_2 = n - 1} f_{l_1,l_2} \ .
\end{equation}
That is, all full grid functions with levels that lie on the bounding diagonal of the level set have a combination coefficient of $\lambda = +1$ while functions with levels on the diagonal below have a coefficient of $\lambda = -1$.
For a general index set, this leads to the formula
\begin{equation}
f^{(\text{s})}_{\vect{l}} = \sum_{\vect{l'} \in \mathcal{I}_\vect{l} } \lambda_{\vect{l'}} f_{\vect{l'}} \ , \qquad \text{with} \quad \lambda_\vect{l'} = \sum_{\vect{z}= \vect{0}}^{\vect{1}} (-1)^{|\vect{z}|_1} \chi^{\mathcal{I}_\vect{l}}(\vect{l'}+\vect{z}) \ ,
\label{eq:combifunct}
\end{equation}
where $\chi$ denotes the indicator function~\cite{harding_adaptive_2016}, defined by the index set $\setI$.

In this paper we use a combination scheme that allows for anisotropic minimal and target levels.
For a more distinctive notation we will denote minimal and target level by $\vect{l}^\text{min}$ and $\vect{l}^\text{max}$, respectively.
The level set has the form
\begin{equation}
\setI_{\vect{l}^\text{min}, \vect{l}^\text{max}} = \left\{ \vect{l'} \in \NN^d : \vect{l'} \in \text{conv}( \vect{l}^\text{min}, \vect{l}^\text{min}+\vect{d}^{(1)}, \dotsc ,  \vect{l}^\text{min}+\vect{d}^{(d)})  \right\} \ ,
\end{equation}
where $\vect{d}^{(i)} = (0,\dotsc, l^\text{max}_i - l^\text{min}_i,\dotsc,0)$, and $\text{conv}$ denoting the convex hull.
This again describes a simplex, but one that is situated at $\vect{l}^\text{min}$ and stretches out to $l^\text{max}_i$ in the $i$-th direction.
The potential for anisotropic minimal and target levels can be used to increase the resolution for more important directions versus directions that may be sampled at lower resolutions, compare \cite{Griebel.Harbrecht:2011}. Additionally, some solvers inherently require a minimal number of grid points, e.g. for the stencils they use, or to make sure that some relevant physical scale can be resolved.

In the case of interpolating a known function, the representations \eqref{eq:sparsefunct} and \eqref{eq:combifunct} are identical.
Thus, the representation error of the combination technique is the same as in \cref{eq:reperror}.
But if the full grid functions constituting the sum in \eqref{eq:combifunct} are derived in a separate way, e.g. by solving a PDE, they carry an additional discretization error and it is not guaranteed that \cref{eq:reperror} holds for the total error.
However, there are cases where it can be shown that it does.
Firstly, if the full grid solutions satisfy an error splitting assumption of the form
\begin{equation}
	(f - f_\vect{l})(x_\vect{l}) = \sum_{i = 1}^d \sum_{\{j_1,\dotsc,j_i\} \subset \{1,\dotsc,d\}} \kappa_{j_1,\dotsc,j_i}(x_\vect{l}; h_{j_1},\dotsc,h_{j_i}) h_{j_1}^2\cdots h_{j_i}^2
\end{equation}
at all full grid points $x_\vect{l}$, with bounded $\abs{\kappa_{j_1,\dotsc,j_i}(x_\vect{l}; h_{j_1},\dotsc,h_{j_i})} \leq K$ and $h_j \coloneqq 2^{-j}$, then the point-wise total error of the combined function has the same asymptotic rate as in \cref{eq:reperror} \cite{Griebel.Schneider.Zenger:1992}.
The same can be shown for the $L^2$-error of the combined function, if the full grid functions are solutions of operator equations with elliptic operators acting on arbitrary Gelfant triples \cite{Griebel.Harbrecht:2014}.
Further analysis for the advection-diffusion equation is found in \cite{Reisinger2007AnalysisOL}, where the same asymptotic convergence rates are shown. Error analysis for the combination technique with multiple time intervals in the case of the advection equation can be found in \cite{Lastdrager01sparse}.

Still, a general theory for arbitrary, particularly nonlinear, PDEs is not available.
For the case of kinetic equations, such as the Vlasov--Poisson system or the gyrokinetic equations, the notion of convergence itself has to be defined first.
While in certain scenarios or temporal regimes, where the nonlinearity of the equations is negligible, a convergence of $f_\vect{l}(t)$ to the exact solution is expected, this is not the case anymore for the long term simulation of turbulent/chaotic behavior.
In this situation, the combination technique can only be expected to work well for the solution of the PDE for very small time frames $\Delta T$~\cite{Lastdrager01sparse}.
In practice, we therefore iterate the application of the combination technique $m$ times in order to arrive at a final time $m \Delta T$.

For the combination step that is necessary after every interval, all subproblems on the different full grids have to be combined in the resulting sparse grid space, according to \cref{eq:combifunct}. For this purpose, the full grid functions are hierarchized and the resulting hierarchical coefficients are added to or subtracted from the sparse grid solution, where the coefficients from missing subspaces are padded with zeros. After that, the sparse grid solution has to be redistributed to the original full grids. There are several ways to do this. Carrying out an $L^2$~projection would yield, on each full grid, the best approximation to the sparse grid function in the $L^2$ norm.
However, in practice, such a projection requires more computational effort because a linear system of equations involving the mass matrix $M_{\vect{l'} \vect{s'},\vect{l} \vect{s}} = \scaprod{\psi_{\vect{l'} \vect{s'}}}{\psi_{\vect{l} \vect{s}}} $ has to be solved. Thus, we use the method of keeping the coefficients of all hierarchical difference spaces contained in the full grid and omitting the rest:
\begin{equation}
\proj_{\vect{l}} f^{(\text{s})} = \proj_{\vect{l}} \left( \sum_{\vect{l'} \in \setI_{\vect{l}^\text{min}, \vect{l}^\text{max}}} \sum_\vect{s} \alpha_{\vect{l'} \vect{s}} \psi_{\vect{l'} \vect{s}} \right) \coloneqq \sum_{ \vect{l}^\text{min} \leq \vect{l'} \leq \vect{l} }  \sum_\vect{s} \alpha_{\vect{l'} \vect{s}} \psi_{\vect{l'} \vect{s}} \ , \qquad \forall \vect{l} \in \setI_{\vect{l}^\text{min}, \vect{l}^\text{max}} \ .
\end{equation}
Afterwards, the dehierarchization operation is applied on each full grid.

For an orthogonal multiscale basis, e.g. orthonormal wavelets, this would be identical to the $L^2$ projection, since in this case $M_{\vect{l'} \vect{s'},\vect{l} \vect{s}} = \delta_{\vect{l'} \vect{s'},\vect{l} \vect{s}}$. For biorthogonal wavelet bases it is not, so the best approximation property is lost. However, as noted earlier, the mass is conserved by this projection when using biorthogonal wavelet bases, since it is wholly contained in the subspace $\lmin$, which is present in all full grids. We argue that, in this instance, mass conservation is the more important property compared to $L^2$ best approximation.

Finally, quantitative analysis of the solution is commonly done by studying certain quantities of interest that are typically averaged over the phase-space.
This is why, in our simulations of the Vlasov-Poisson system, we assess the quality of the combined solution by analyzing its performance in two benchmark problems via the time traces of the potential energy.
A similar approach was taken in \cite{lago_exahd_2020} for simulations of the gyrokinetic equations with the code GENE~\cite{jenko_global_2013}.
A full convergence study of the combination technique failed, however, since for many index sets the simulations became unstable.
Therefore, the main goal of this paper is to show that the use of biorthogonal wavelet bases helps to alleviate this problem.

\section{Mass-conserving solvers for hyperbolic problems}\label{sec:solver}

Naturally, the total mass can only be conserved in a combination technique with the novel hierarchical bases if the underlying solvers are also mass-conserving.
We therefore choose a finite volume advection solver and a semi-Lagrangian solver for the kinetic equations that both conserve mass and, moreover, operate on a structured six-dimensional grid.

By contrast, plasma turbulence codes such as \gene that are closely inspired by predictions for experimental nuclear fusion devices typically incorporate optimizations and simplifications that make a numerical analysis on the system level difficult.  
Moreover, they do not conserve mass or other momenta, since they only simulate a perturbation from equilibrium instead of the full distribution function.

\subsection{Finite volume advection solver}\label{subsec:solver:advection}
As a first benchmark problem, we consider the incompressible advection equation
\begin{equation}\label{eq:solver:advection:incompressible}
  \deriv{u(\vecx, t)}{t} + \vect{a} \cdot \grad_{\vecx} u(\vecx, t) = 0
\end{equation}
on the unit hypercube $\Omega = [0,1]^d$ in $d$ dimensions.
The advection velocity $\vect{a}$ is $\vect{\mathbf{1}}$, constant throughout $\Omega$.
Note that a purely diagonal advection is the most challenging direction for combination technique simulations.

The spatial boundary conditions are periodic, which makes this problem well-suited for our purposes, since the analytical solution can be obtained in a straightforward manner by periodically translating the initial distribution.
This makes a direct Monte Carlo integration of the error possible, \cf \cref{subsec:experiments:advection}.

A straightforward discretization is the first-order finite volume upwinding scheme for positive velocities
\begin{equation}\label{eq:solver:advection:stencil}
  \begin{aligned}
    \grad_i u = \deriv{u}{x_i} & \approx \frac{u(\vecx, t-\Delta t) - u(\vecx - \Delta x_i, t-\Delta t)}{\Delta x_i}, \\
    \deriv{u}{t}               & \approx \frac{u(\vecx, t) - u(\vecx, t-\Delta t)}{\Delta t},
  \end{aligned}
\end{equation}
which conserves the mass \Lone{u}---assuming positivity of $u$---up to machine precision.
Note that the numerical value of the mass (at initial time) can vary slightly depending on the resolution due to the numerical interpolation of the initial solution.

Plugging \cref{eq:solver:advection:stencil} into \cref{eq:solver:advection:incompressible} and solving for $u(\vect{x}, t)$ results in the explicit Euler time-stepping scheme
\begin{equation}
  u(\vecx,t) = u(\vecx,t-\Delta t) - \grad u \cdot \vect{a} \cdot \Delta t,
\end{equation}
which is first-order accurate in both space and time (provided that the loose CFL conditions of the combination technique are fulfilled, \cf~\cite{bellebaum_evaluation_2018}).
The solver is implemented directly into the distributed combination technique framework \discotec.

\subsection{Semi-Lagrangian Vlasov plasma solver}\label{subsec:solver:selalib}

In a second example, we study a simulation of the Vlasov--Poisson system in 3d3v (three spatial and three velocity dimensions).
We focus on the case of electrons in a neutralizing background. Let $f(\vect{x},\vect{v},t)$ be the distribution function of the electron species of a plasma in phase space $(\vect{x},\vect{v}) \in \Omega \times \R^3$, $\Omega \subset \R^3$. This distribution function evolves according to the Vlasov equation
\begin{equation}
	\partial_t f(\vect{x},\vect{v},t) + \vect{v} \cdot \nabla_\vect{x} f(\vect{x},\vect{v},t) + \frac{q}{m} \left( \vect{E}(\vect{x},t) + \vect{v}\times \vect{B}(\vect{x},t) \right) \cdot \nabla_\vect{v} f(\vect{x},\vect{v},t) = 0,
\end{equation}
where $q$ and $m$ denote electron charge and mass, respectively, $\vect{E}$ the electric and $\vect{B}$ the magnetic field.
The self-consistent fields are given by Maxwell's equations or---for low-frequency phenomena---by Poisson's equation.
In this paper, we focus on the Poisson equation given by
\begin{equation}
	- \Delta_\vect{x}  \phi(\vect{x},t) = 1 - \int f(\vect{x},\vect{v},t) \, \text{d} \vect{v}, \quad \vect{E}(\vect{x},t) = - \nabla_\vect{x} \phi(\vect{x},t).
\end{equation}
and $\vect{B}(\vect{x},t) = 0$.

Several discretization schemes are possible in order to approximate the Vlasov equation: methods of Eulerian, semi-Lagrangian, and Lagrangian type.
While Lagrangian particle methods overcome the curse of dimensionality to some extent, they suffer from inherent numerical noise.
In this paper, instead, we propose to use the sparse grid combination technique applied to a grid-based method in order to overcome the curse of dimensionality with a noise-free method.
We choose the semi-Lagrangian method as the basis for our full grid simulations, since semi-Lagrangian schemes allow for larger time steps in explicit time propagation schemes compared to Eulerian methods.

The semi-Lagrangian method discretizes the distribution function on a grid and the propagation is based on the characteristic equations
\begin{equation}
	\frac{\text{d} \vect{X}}{\text{d} t} = \vect{V}, \quad \frac{\text{d} \vect{V}}{\text{d} t} = \frac{q}{m} \left(\vect{E} + \vect{V} \times \vect{B}\right)
\end{equation}
along which the distribution function is constant, i.e.
$$f(\vect{x},\vect{v},t) = f_0(\vect{X}(0;\vect{x},\vect{v},t), \vect{V}(0;\vect{x},\vect{v},t)),$$ where $(\vect{X}(\tau;\vect{x},\vect{v},t), \vect{V}(\tau;\vect{x},\vect{v},t))$ is the solution of the characteristic curve at time $\tau$ starting at $(\vect{x},\vect{v})$ at time $t$ and $f_0$ is the initial solution at time $\tau=0$.

Since the fields depend on the distribution function, we cannot follow the characteristics from final to initial time.
However, over small time intervals, approximating the fields by a constant value in time is a good approximation.
Based on this assumption, the Vlasov equation can be solved in two steps:
Solving the ODE system of the characteristics is followed by evaluating the interpolated distribution function at the foot of the characteristic.
Using an operator splitting method and solving the $\vect{x}$ and the $\vect{v}$ advection separately, the advection coefficients are constant in each time step so that the solution of the system of ODEs can be found analytically.
This algorithm has been proposed in \cite{cheng_integration_1976}.
Several schemes are conceivable for the one-dimensional interpolations.
In our implementation, we use Lagrangian interpolation, since the interpolation is local, a fact which is favorable in the context of distributed domains.
It can be shown that this scheme conserves both mass and momentum up to round-off errors.

For our simulations, we use the semi-Lagrangian method implemented for high performance computing simulations in SeLaLib \citep{kormann_massively_2019}.
In order to couple \selalib to the \discotec framework, some adaptations were implemented regarding computation and communication of the distribution function.
Fortunately, the coordinate placement used in \selalib naturally results in nested coordinates when doubling the resolution, which matches the requirements of the combination technique perfectly.

\section{Combination technique experiments}\label{sec:experiments}
In our combination technique simulation experiments, the solvers described in \cref{sec:solver} are used to perform a time step update on each of the component grids.
These solver updates are independent between the component grids and can be performed in parallel.
Information between the component grids is only exchanged in the combination step:
The recombination transforms the data on each component grid into their hierarchical representations---coefficients corresponding to either of $\hbasishat$, $\hbasisbiorthogonal$, $\hbasisfullweighting$---and sums them up to a common sparse grid according to the combination formula (\cref{eq:combifunct}).
The decombination then \enquote{scatters}, \ie projects, the sparse grid data back to the component grids and, afterwards, performs the inverse transformation on each grid to recover the interpolating coefficients.
Then, the next time step can begin.

The three numerical experiments were designed to investigate different behaviors:
First, to verify that mass is now conserved up to machine precision; second, to showcase the error properties of the combination technique using the different basis functions w.r.t.~an analytical solution, \cf \cref{subsec:experiments:advection}; and, third, to apply the different combination technique approaches to plasma physics scenarios while comparing them to the monolithic full grid solution at a fixed amount of available memory, \cf \cref{subsec:experiments:landau,subsec:experiments:instability}.

In order to achieve a fair comparison, the number of degrees of freedom for the full grid simulations is calculated as the number of points on that single grid
\begin{equation}\label{eqn:experiments:num_fg_dof}
  \#\si{\dof}_{\text{fg}} = \sum_{i=1}^{d} 2^{\scalarl^\text{fg}_i},
\end{equation}
whereas for the combination technique, we need to consider the sum of points on all component grids
\begin{equation}\label{eqn:experiments:num_combi_dof}
  \#\si{\dof}_\text{ct} = \sum_{\justl \in \mathcal{I} | \lambda_\justl \neq 0}   \sum_{i=1}^{d} 2^{\scalarl_i}.
\end{equation}

This implies that the combination technique solution is not necessarily \enquote{cheaper} than a full grid of level \lmax, and that the index set $\mathcal{I}$ needs to be chosen carefully.
We use only double-precision floats for the representation of the (interpolating and hierarchical) basis coefficients, such that the memory consumption can be obtained as ${\# \si{\dof} \cdot 8 / 2^{30}\, \si{\gibi\byte}}$.

Another aspect that is common to all experiments is the choice of the recombination interval $\Delta T$.
\citeauthor{Lastdrager01sparse}~\cite{Lastdrager01sparse} analyzed the different error terms introduced by the combination technique with hat functions for advection-type problems.
They could show that by recombination at constant simulation time intervals throughout the simulation, the leading time-error term would be of first order~\cite[Eq.~(32)]{Lastdrager01sparse}.
However, they did not take the solver's own time-dependent error into account, but only the solver's spatial error.
Hence, we restrict ourselves to recombination after each solver time step $\Delta t$ for our experiments, \ie $\Delta T = \Delta t$, in order to avoid unforeseen numerical artifacts originating from different time ranges and error expansions between the solver and the combination technique.

In the simulations, the mass (and other properties) of the combination technique solution are obtained by evaluating the same property on each component grid and summing with the combination formula
\begin{equation}\label{eqn:advection:ctnorm}
  Q_{\text{ct}}(f) \coloneqq \sum_{\vect{l} \in \setI} \lambda_{\vect{l}} \cdot Q(f_\vect{l}).
\end{equation}

Note that the formula is only equivalent to the interpolated sparse grid properties
\begin{equation}\label{eqn:advection:sgnorm}
Q_{\text{sg}}(f) \coloneqq Q\Bigg(\sum_{\vect{l} \in \setI} \lambda_{\vect{l}} \cdot f_\vect{l}\Bigg)
\end{equation}%
for linear operators, e.g.~the mass.
For instance in the maximum norm, one can easily imagine that the combined maximum is not the same as the sparse grid maximum, since the maximal function value will usually be attained at different points in the component grids---and the sparse grid point with maximal value will likely not be present in each of the component grids.

\Cref{subsec:experiments:landau,subsec:experiments:instability} are going to deal with both the (linear) mass and the (nonlinear) potential energy $\phi(f)$ as combined quantities $Q_\text{ct}(f)$.
We validated the approach of using the combined potential energy with coarsely resolved CT plasma simulations, where both $\phi_\text{ct}(f)$ and the interpolated $\phi_\text{sg}(f)$ could be feasibly computed.
The result was that, for those time intervals where the simulation was stable, the properties would be in close agreement.
We can therefore assume that the combination technique linearization assumption
\begin{equation}\label{eq:linearization}
  Q_{\text{sg}}(f) \approx Q_{\text{ct}}(f)
\end{equation}
for a general quantity of interest $Q(\cdot)$ holds sufficiently in the cases that are physically relevant.
And conversely, we observed that physical bounds---such as the positivity of the potential energy---are only violated at the onset of numerical instabilities.

The following sections use the combined quantities of interest unless otherwise indicated.

\subsection{Advecting a Gaussian}\label{subsec:experiments:advection}

The finite volume advection solver introduced in \cref{subsec:solver:advection} is used to compare the different hierarchical basis functions on an analytically solvable scenario in arbitrary dimensions.

The unknown concentration $u$ is initialized at $t=0$ by a Gaussian distribution
\begin{equation}\label{eq:solver:advection:initial}
  u(\vecx, t=0) = \exp\left({-\sum_{i=1}^d \left(x_i-\frac{1}{2}\right)^2 \cdot \frac{1}{\sigma^2}}\right),
\end{equation}
that is, a multivariate normal distribution with standard deviation $\sigma = \frac{1}{3}$.
It is normalized, such that the maximum value is always $1$, independently of $d$.
Gaussians are the typical initial conditions in velocity space for plasma physics simulations, and they can be particularly challenging to be represented on sparse grids~\cite{Pflueger10Spatially}.

The solver time step size $\dt$ is set to \num{1e-4}, to fulfill the CFL condition for all considered resolutions, and recombination takes place after each time step.
While the full grid reference simulations are each run on a single grid at different isotropic resolutions, all combination technique simulations have a fixed minimum level of $\lmin = (2)^d$ but different target levels $\lmax$, which correspond to the resolutions of the reference simulations.
This results in an increasing number of component grids as $\lmax$ increases.

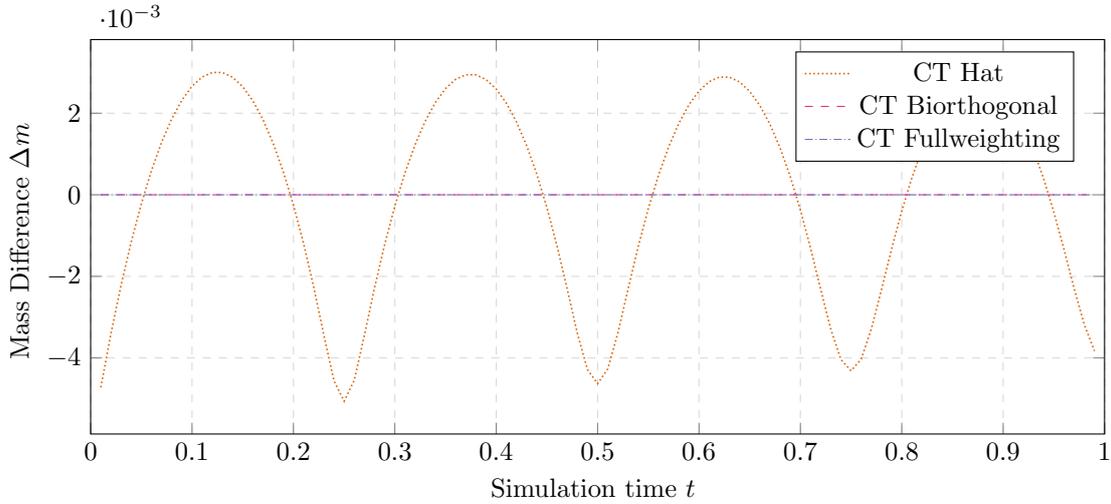
\begin{figure}[H]
    \begin{tikzpicture}[remember picture]
      \begin{axis}[
          timeplot,
          height=0.3\textheight,
          ylabel=Mass Difference $ \Delta m $, 
          table/y index=2,
          table/y expr=\xintthefloatexpr{\thisrowno{2}-0.18596629200709755}\relax,
          xmin=0,
          xmax=1,
        ]
        \addplot[hat]
        table[] {data/advection/advection_norms_hat_3D_2-11.csv};
        \addplot[biorthogonal]
        table[] {data/advection/advection_norms_biorthogonal_periodic_3D_2-11.csv};
        \addplot[fullweighting]
        table[] {data/advection/advection_norms_fullweighting_periodic_3D_2-11.csv};
        \legend{CT Hat, CT Biorthogonal, CT Fullweighting}
      \end{axis}
    \end{tikzpicture}
    \caption{Difference $ \Delta m $ between the combined masses $\sum_{\vect{l} \in \setI} \lambda_{\vect{l}} \cdot \int {u_\vect{l}(\vecx, t)} \,d\vecx$ and analytical mass $\int {\uana(\vecx, \cdot)} \,d\vecx \approx \num[round-mode=figures,round-precision=3]{0.18596629200709755}$ for a combination technique 3D advection scenario with $\lmin = (2, \,2, \,2)$, $\lmax = (11, \,11, \,11)$ (136 component grids). Using hierarchical hat functions leads to fluctuations of up to \SI{2.7}{\percent} of the analytical mass. The maxima in the mass are reached whenever the Gaussian's maximum passes through those regions of the sparse grid where there are fewer degrees of freedom. By comparison, the biorthogonal and full weighting bases conserve the mass in the simulation up to an accuracy of \num{1e-10}.}
  \label{fig:advection:onenorm}
\end{figure}

We first investigate the combination technique's behavior with respect to the one known invariant of the solver:
$\uana$ is chosen such that the mass \Lone{\uana}---assuming positivity of $u$---is approximately $0.570792^d$, \cf \cref{eq:solver:advection:initial}.
The full grid simulations all conserve the mass up to machine precision.
\Cref{fig:advection:onenorm} shows the mass fluctuations throughout the simulation time in the combination technique solutions.
It clearly states that the $\hbasisbiorthogonal$ and $\hbasisfullweighting$ bases lead to the desired conservation of mass, in contrast to $\hbasishat$.

Next, we consider different errors at $t=1.0$, \ie after one full cycle of diagonally advecting the initial solution through the domain.
Since the true solution is known analytically, the relative error norms w.r.t.~the analytical solution are obtained by Monte Carlo integration
\begin{equation}
  \begin{aligned}
    \frac{\norm{\uct - \uana}_1}{\norm{\uana}_1}(t) & \approx \frac{\sum_{j=1}^N \abs{\uct(\vecx_j,t)-\uana(\vecx_j,t)}}{\sum_{j=1}^N \abs{\uana(\vecx_j,t)}},  \\
    \frac{\norm{\uct - \uana}_2}{\norm{\uana}_2}(t) & \approx \frac{\sum_{j=1}^N \abs{\uct(\vecx_j,t)-\uana(\vecx_j,t)}^2}{\sum_{j=1}^N \abs{\uana(\vecx_j,t)}^2}, \\
    \frac{\norm{\uct - \uana}_\infty}{\norm{\uana}_\infty}(t) & \approx \frac{\max_{\vecx_j}\abs{\uct(\vecx_j,t)-\uana(\vecx_j,t)}}{\max_{\vecx_j}\abs{\uana(\vecx_j,t)}},
  \end{aligned}
\end{equation}
on $N=\num{1e5}$ randomly sampled coordinates $\vecx_j$ (the pseudorandom numbers are generated using the Mersenne Twister algorithm).
$N$ is chosen numerically such that the variance of the Monte Carlo integrals over multiple runs with different random coordinates is sufficiently small.
Using random coordinates---not sparse grid collocation points---is particularly important since the analytical solution is not contained in the linear ansatz space;
it is necessary to capture the error between collocation points especially for low-resolution simulations.

\Cref{fig:advection:errors} displays several characteristics of the different solutions: For the same \lmax, the combination technique solutions require drastically smaller numbers of degrees of freedom than the full grid solutions, while providing similar numerical accuracy in all considered error norms.
Furthermore, using the mass-conserving basis functions reduces the errors further, an effect that gets more significant as the dimensionality increases.

\begin{figure}[p!]
  \centering
  \begin{tikzpicture}
    \begin{groupplot}[group style={group name=advError,group size= 1 by 3,vertical sep=1.5cm },ymax=1.3, height=0.3\textheight, width=0.6\textwidth]
      \nextgroupplot[table/x=dof, table/y=L2error, table/col sep=comma,ylabel={$\frac{\norm{\uct - \uana}_2}{\norm{\uana}_2}(t=1)$},xmode=log, ymode=log, filter discard warning=false]
      \addplot[fg, twodim, discard if not={dim}{2}, discard if not={basis}{fg}]
      table[] {data/advection_conservation.csv} coordinate (last2fg) {};
      \addplot[hat, twodim, discard if not={dim}{2}, discard if not={basis}{hat}]
      table[] {data/advection_conservation.csv} coordinate (last2hat) {};
      \addplot[fullweighting, twodim, discard if not={dim}{2}, discard if not={basis}{fullweighting_periodic}]
      table[] {data/advection_conservation.csv} coordinate (last2fw) {};
      \addplot[biorthogonal, twodim, discard if not={dim}{2}, discard if not={basis}{biorthogonal_periodic}]
      table[] {data/advection_conservation.csv} coordinate (last2bio) {};
      \node[draw,red,ellipse,inner sep=1ex,fit=(last2fg)(last2hat)(last2fw)(last2bio)] {};
      \addplot[fg, fourdim, discard if not={dim}{4}, discard if not={basis}{fg}]
      table[]{data/advection_conservation.csv};
      \addplot[hat, fourdim, discard if not={dim}{4}, discard if not={basis}{hat}]
      table[]{data/advection_conservation.csv};
      \addplot[fullweighting, fourdim, discard if not={dim}{4}, discard if not={basis}{fullweighting_periodic}]
      table[]{data/advection_conservation.csv};
      \addplot[biorthogonal, fourdim, discard if not={dim}{4}, discard if not={basis}{biorthogonal_periodic}]
      table[]{data/advection_conservation.csv};
      \addplot[fg, sixdim, discard if not={dim}{6}, discard if not={basis}{fg}]
      table[]{data/advection_conservation.csv};
      \addplot[hat, sixdim, discard if not={dim}{6}, discard if not={basis}{hat}]
      table[]{data/advection_conservation.csv};
      \addplot[fullweighting, sixdim, discard if not={dim}{6}, discard if not={basis}{fullweighting_periodic}]
      table[]{data/advection_conservation.csv};
      \addplot[biorthogonal, sixdim, discard if not={dim}{6}, discard if not={basis}{biorthogonal_periodic}]
      table[]{data/advection_conservation.csv};
      \addplot[fg,black] coordinates {(1e4,1e2)};\label{plot:fg}
      \addplot[hat,black] coordinates {(1e4,1e2)};\label{plot:hat}
      \addplot[fullweighting,black] coordinates {(1e4,1e2)};\label{plot:fullweighting}
      \addplot[biorthogonal,black] coordinates {(1e4,1e2)};\label{plot:biorthogonal}
      \addplot[twodim,] coordinates {(1e4,1e2)};\label{plot:twodim}
      \addplot[fourdim,] coordinates {(1e4,1e2)};\label{plot:fourdim}
      \addplot[sixdim,] coordinates {(1e4,1e2)};\label{plot:sixdim}
      \coordinate (top) at (rel axis cs:0,1);

      \nextgroupplot[table/x=dof, table/y=L1error, table/col sep=comma, title={},ylabel={$\frac{\norm{\uct - \uana}_1}{\norm{\uana}_1}(t=1)$},xmode=log, ymode=log, filter discard warning=false]
      \addplot[fg, twodim, discard if not={dim}{2}, discard if not={basis}{fg}]
      table[] {data/advection_conservation.csv} node (last1fg) {};
      \addplot[hat, twodim, discard if not={dim}{2}, discard if not={basis}{hat}]
      table[] {data/advection_conservation.csv} node (last1hat) {};
      \addplot[fullweighting, twodim, discard if not={dim}{2}, discard if not={basis}{fullweighting_periodic}]
      table[] {data/advection_conservation.csv} node (last1fw) {};
      \addplot[biorthogonal, twodim, discard if not={dim}{2}, discard if not={basis}{biorthogonal_periodic}]
      table[] {data/advection_conservation.csv} node (last1bio) {};
      \node[draw,red,ellipse,inner sep=0.7ex,fit=(last1fg)(last1hat)(last1fw)(last1bio)] {};
      \addplot[fg, fourdim, discard if not={dim}{4}, discard if not={basis}{fg}]
      table[]{data/advection_conservation.csv};
      \addplot[hat, fourdim, discard if not={dim}{4}, discard if not={basis}{hat}]
      table[]{data/advection_conservation.csv};
      \addplot[fullweighting, fourdim, discard if not={dim}{4}, discard if not={basis}{fullweighting_periodic}]
      table[]{data/advection_conservation.csv};
      \addplot[biorthogonal, fourdim, discard if not={dim}{4}, discard if not={basis}{biorthogonal_periodic}]
      table[]{data/advection_conservation.csv};
      \addplot[fg, sixdim, discard if not={dim}{6}, discard if not={basis}{fg}]
      table[]{data/advection_conservation.csv};
      \addplot[hat, sixdim, discard if not={dim}{6}, discard if not={basis}{hat}]
      table[]{data/advection_conservation.csv};
      \addplot[fullweighting, sixdim, discard if not={dim}{6}, discard if not={basis}{fullweighting_periodic}]
      table[]{data/advection_conservation.csv};
      \addplot[biorthogonal, sixdim, discard if not={dim}{6}, discard if not={basis}{biorthogonal_periodic}]
      table[]{data/advection_conservation.csv};
      \nextgroupplot[table/x=dof, table/y=L0error, table/col sep=comma, title={},ylabel={$\frac{\norm{\uct - \uana}_\infty}{\norm{\uana}_\infty}(t=1)$},xlabel = {\#\si{\dof}},xmode=log, ymode=log, filter discard warning=false]
      \addplot[fg, twodim, discard if not={dim}{2}, discard if not={basis}{fg}]
      table[] {data/advection_conservation.csv} node (last0fg) {};
      \addplot[hat, twodim, discard if not={dim}{2}, discard if not={basis}{hat}]
      table[] {data/advection_conservation.csv} node (last0hat) {};
      \addplot[fullweighting, twodim, discard if not={dim}{2}, discard if not={basis}{fullweighting_periodic}]
      table[] {data/advection_conservation.csv} node (last0fw) {};
      \addplot[biorthogonal, twodim, discard if not={dim}{2}, discard if not={basis}{biorthogonal_periodic}]
      table[] {data/advection_conservation.csv} node (last0bio) {};
      \node[draw,red,ellipse,inner sep=0ex,fit=(last0fg)(last0hat)(last0fw)(last0bio)] {};
      \addplot[fg, fourdim, discard if not={dim}{4}, discard if not={basis}{fg}]
      table[]{data/advection_conservation.csv};
      \addplot[hat, fourdim, discard if not={dim}{4}, discard if not={basis}{hat}]
      table[]{data/advection_conservation.csv};
      \addplot[fullweighting, fourdim, discard if not={dim}{4}, discard if not={basis}{fullweighting_periodic}]
      table[]{data/advection_conservation.csv};
      \addplot[biorthogonal, fourdim, discard if not={dim}{4}, discard if not={basis}{biorthogonal_periodic}]
      table[]{data/advection_conservation.csv};
      \addplot[fg, sixdim, discard if not={dim}{6}, discard if not={basis}{fg}]
      table[]{data/advection_conservation.csv};
      \addplot[hat, sixdim, discard if not={dim}{6}, discard if not={basis}{hat}]
      table[]{data/advection_conservation.csv};
      \addplot[fullweighting, sixdim, discard if not={dim}{6}, discard if not={basis}{fullweighting_periodic}]
      table[]{data/advection_conservation.csv};
      \addplot[biorthogonal, sixdim, discard if not={dim}{6}, discard if not={basis}{biorthogonal_periodic}]
      table[]{data/advection_conservation.csv};
      \coordinate (bot) at (rel axis cs:1,0);
    \end{groupplot}
    \coordinate[xshift=10ex](legendpos) at (advError c1r1.south east);
    \matrix[
      matrix of nodes,
      anchor=west,
      draw,
      inner sep=0.2em
    ]at(legendpos)
    {
      \ref{plot:fg} & Full Grid                     & [5pt]\\
      \ref{plot:hat} & CT $\hbasishat$                  & [5pt]\\
      \ref{plot:fullweighting} & CT $\hbasisfullweighting$                & [5pt]\\
      \ref{plot:biorthogonal} & CT $\hbasisbiorthogonal$                  & [10pt]\\\par\bigskip
      \ref{plot:twodim} & 2D                          & [5pt]\\
      \ref{plot:fourdim} & 4D                          & [5pt]\\
      \ref{plot:sixdim} & 6D \\};
  \end{tikzpicture}
  \caption{Monte Carlo errors for the advection solver over the number of degrees of freedom used.
  Curves are shown for dimensionalities of 2, 4, and 6, comparing the full grid reference solutions to combination technique solutions with different hierarchical basis functions at $t=1$.
  For the comparison in the red ellipses (2D, $\lmin=(2,\,2)$, $\lmax=(11,\,11)$), the $u$ fields are plotted in \cref{fig:advection:difference} at $t=0.5$}\,.
  \label{fig:advection:errors}
\end{figure}
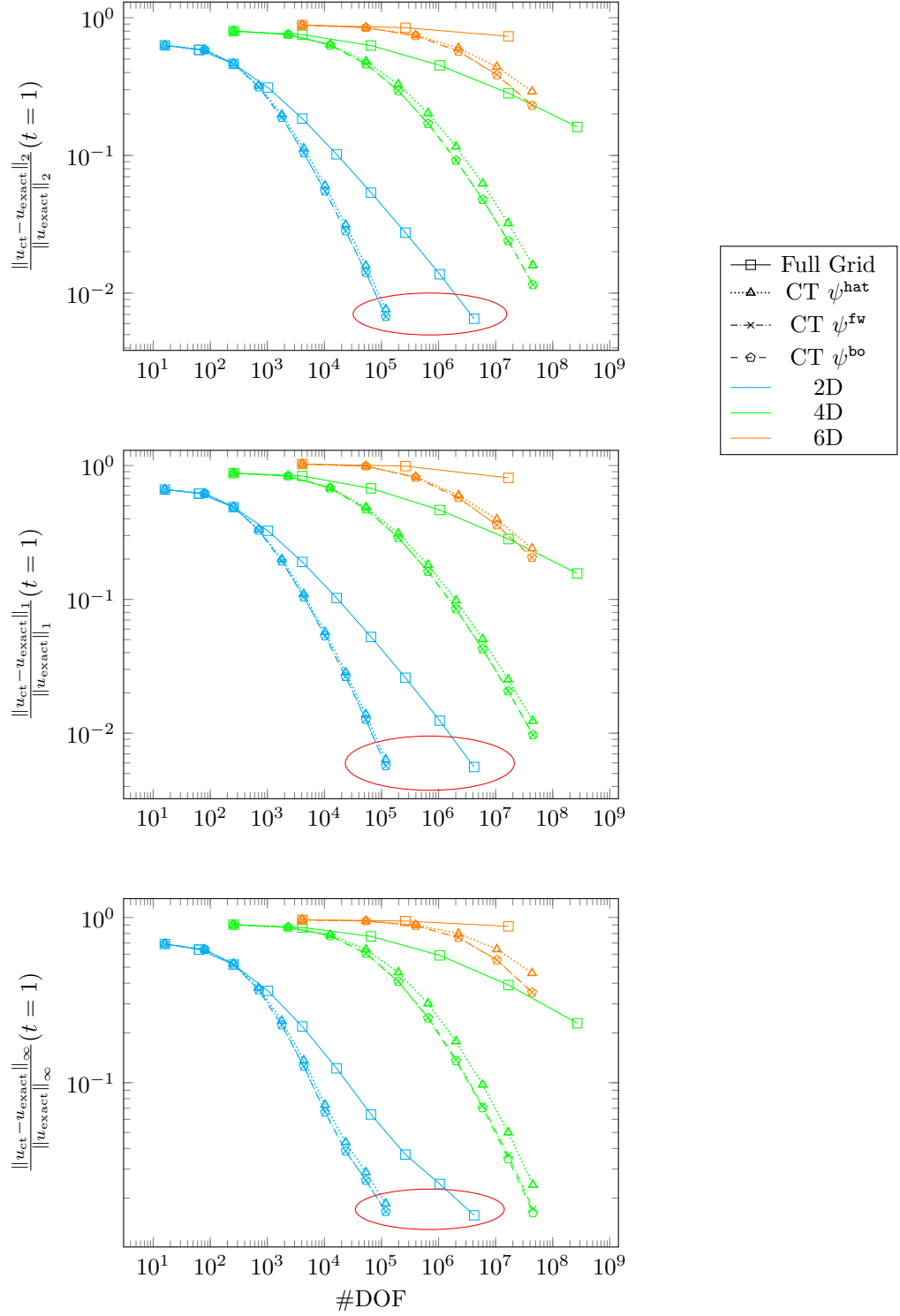
\begin{figure}[p!]
  \centering
  \hspace*{8ex}
  \includegraphics[width=0.3\textwidth]{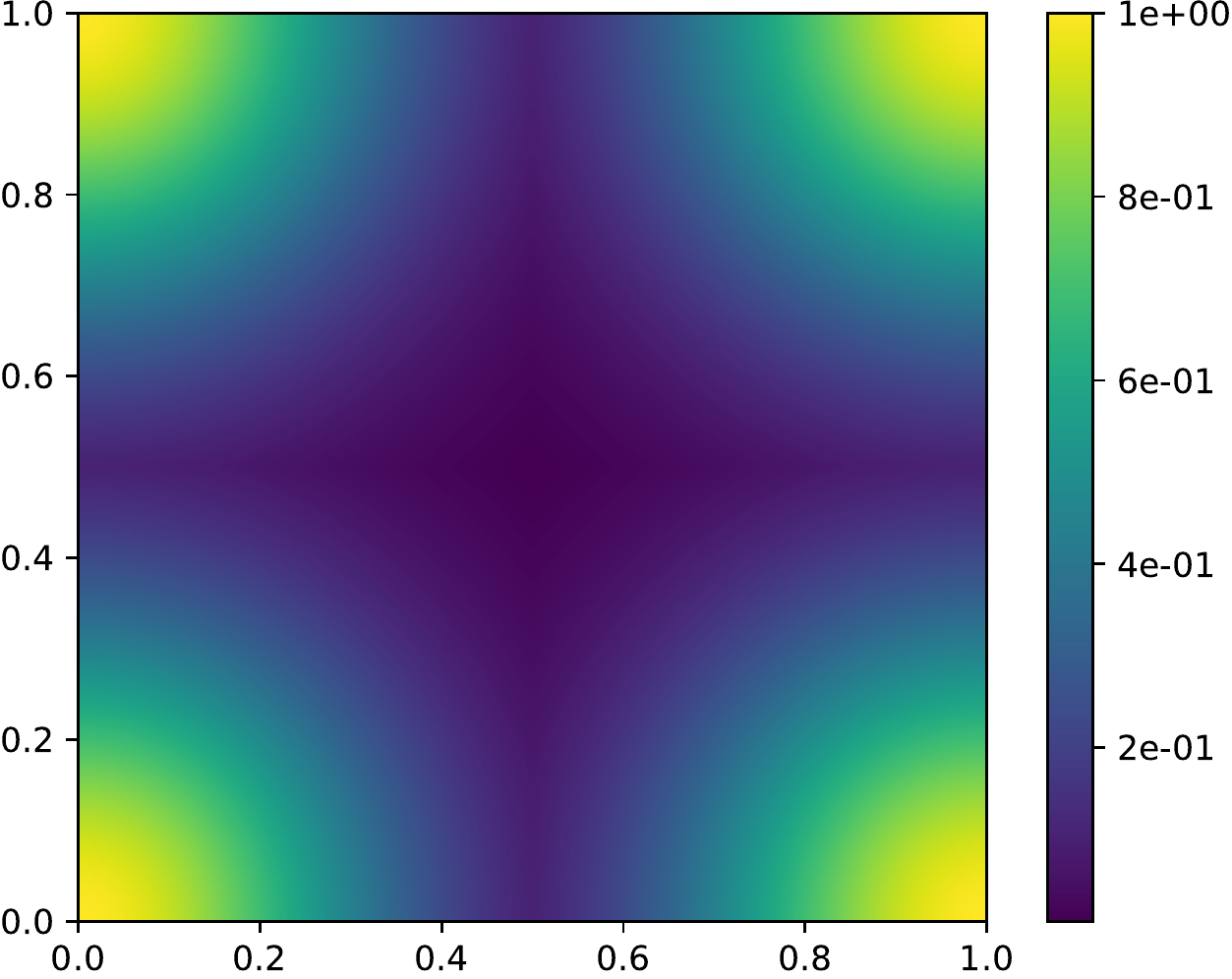}
  \caption{Analytical solution $\uana$ at time $t=0.5$.
  The Gaussian has moved along half of the diagonal, such that the maximum is attained exactly at the corners of the domain.}
  \label{fig:advection:analytical}
  \begin{subfigure}[]{0.08\textwidth}
    \includegraphics[width=\textwidth]{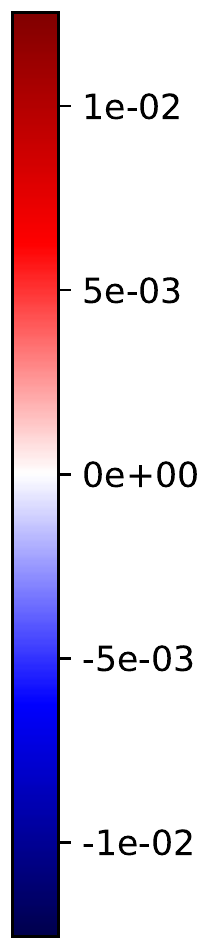}
  \end{subfigure}\hfill
  \begin{subfigure}[]{0.2\textwidth}
    \centering
    \begin{subfigure}[]{\textwidth}
      \centering
      \begin{tikzpicture}[remember picture]
        \node (n1) at (0,0) {
          \includegraphics[width=\linewidth]{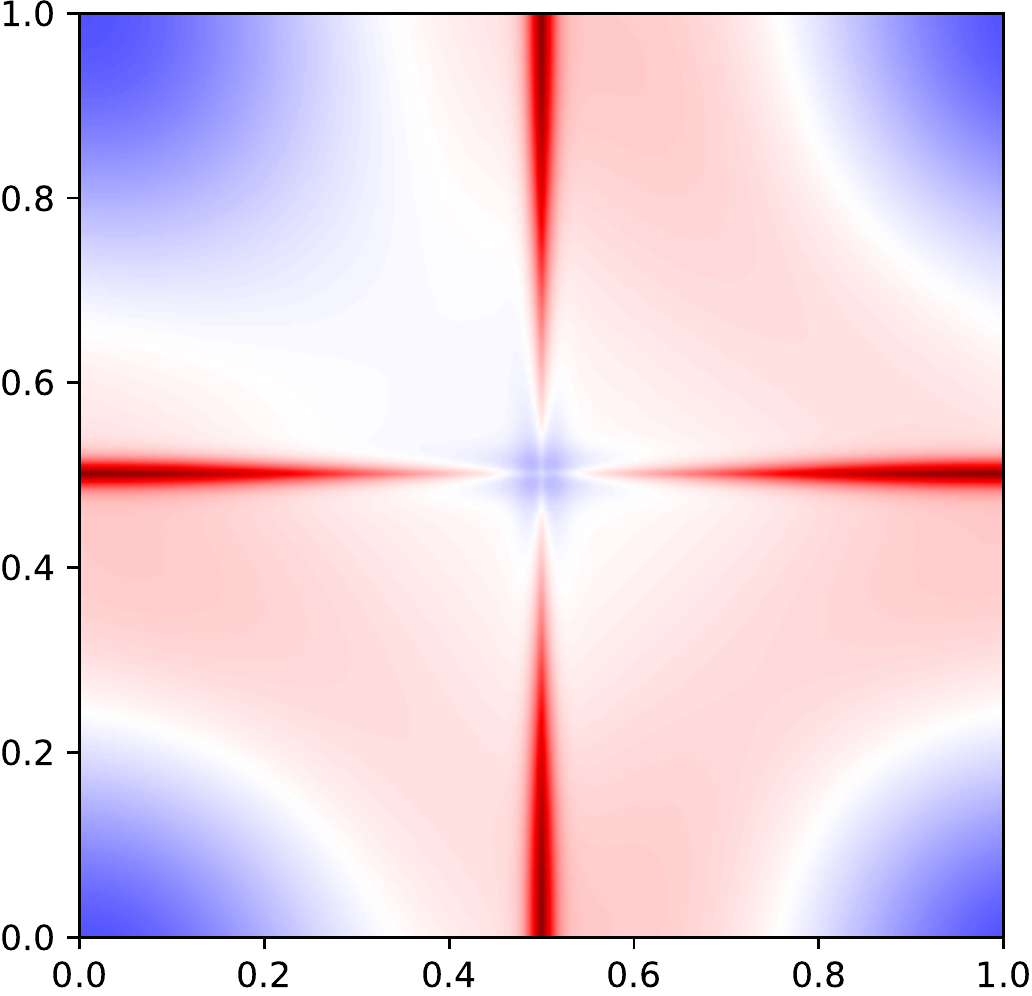}
        };
      \end{tikzpicture}
      \caption*{hierarchical hat basis \hbasishat}
      \label{subfig:advection:ct_hat_error}
    \vspace*{1em}
    \end{subfigure}
    \begin{subfigure}[]{\textwidth}
      \centering
      \begin{tikzpicture}[remember picture]
        \node (n1) at (0,0) {
          \includegraphics[width=\linewidth]{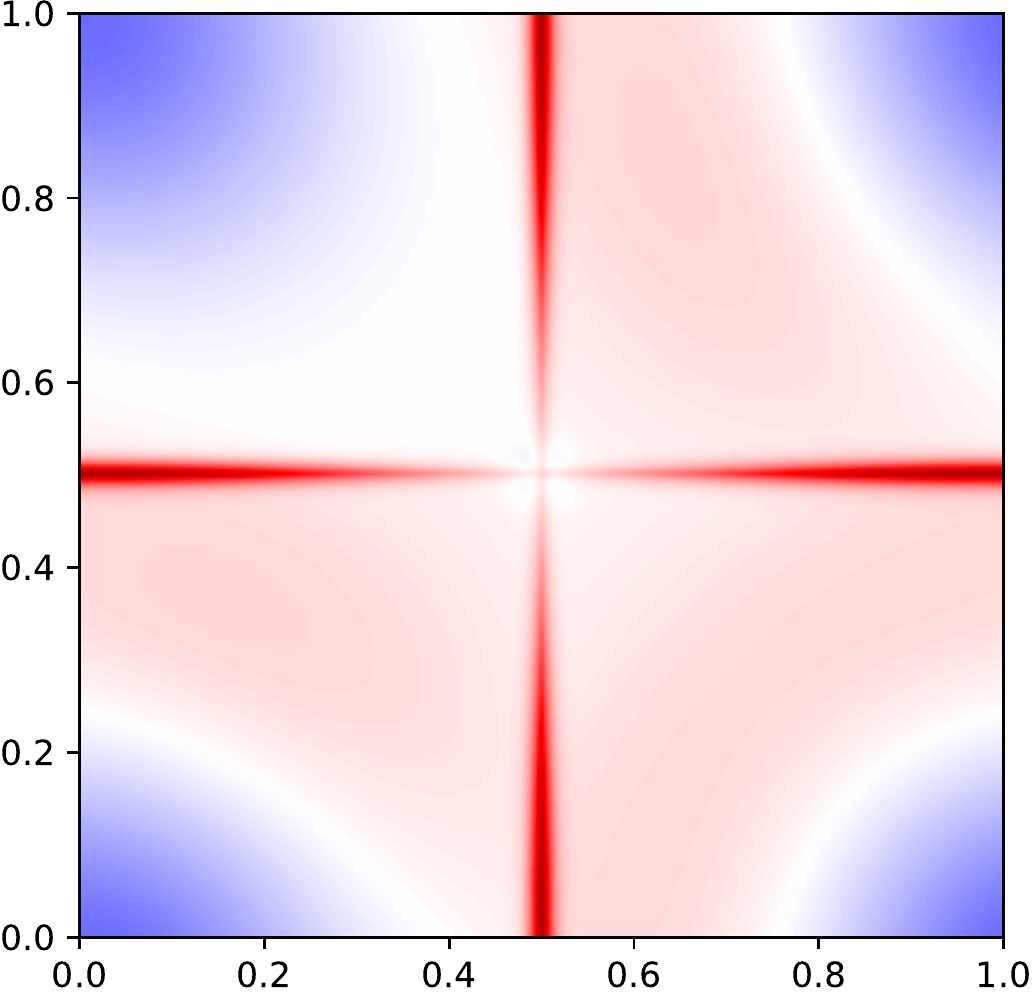}
        };
      \end{tikzpicture}
      \caption*{biorthogonal basis \hbasisbiorthogonal}
      \label{subfig:advection:ct_biorthogonal_error}
    \vspace*{1em}
    \end{subfigure}
    \begin{subfigure}[]{\textwidth}
      \centering
      \begin{tikzpicture}[remember picture]
        \node (n1) at (0,0) {
          \includegraphics[width=\linewidth]{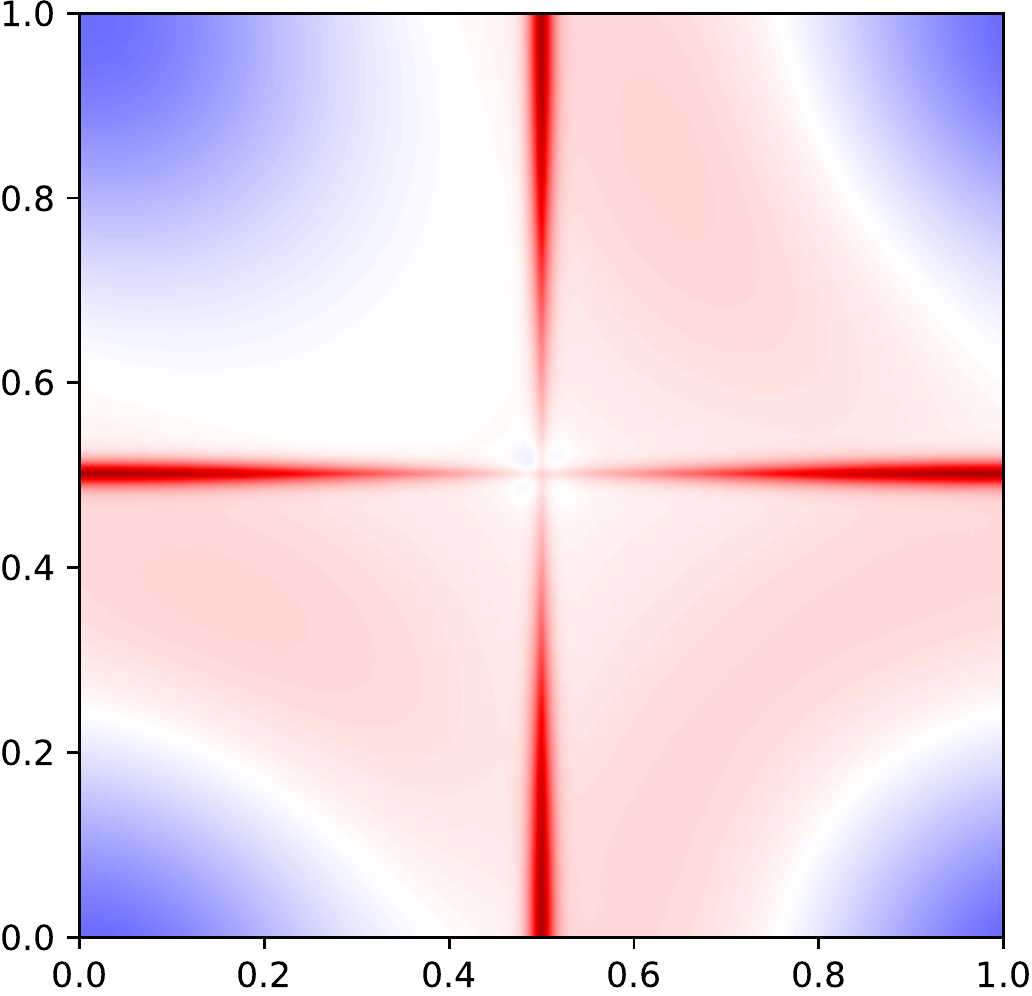}
        };
      \end{tikzpicture}
      \caption*{full weighting basis \hbasisfullweighting}
      \label{subfig:advection:ct_fullweighting_error}
    \end{subfigure}
    \caption{Total combination technique errors ${\uct - \uana}$}
    \label{subfig:advection:ct_error}
  \end{subfigure}%
  \hfill$=$\hfill
  \begin{subfigure}[]{0.2\textwidth}
    \centering
    \vspace*{11.45em}
    \begin{tikzpicture}[remember picture]
      \node (n1) at (0,0) {
        \includegraphics[width=\linewidth]{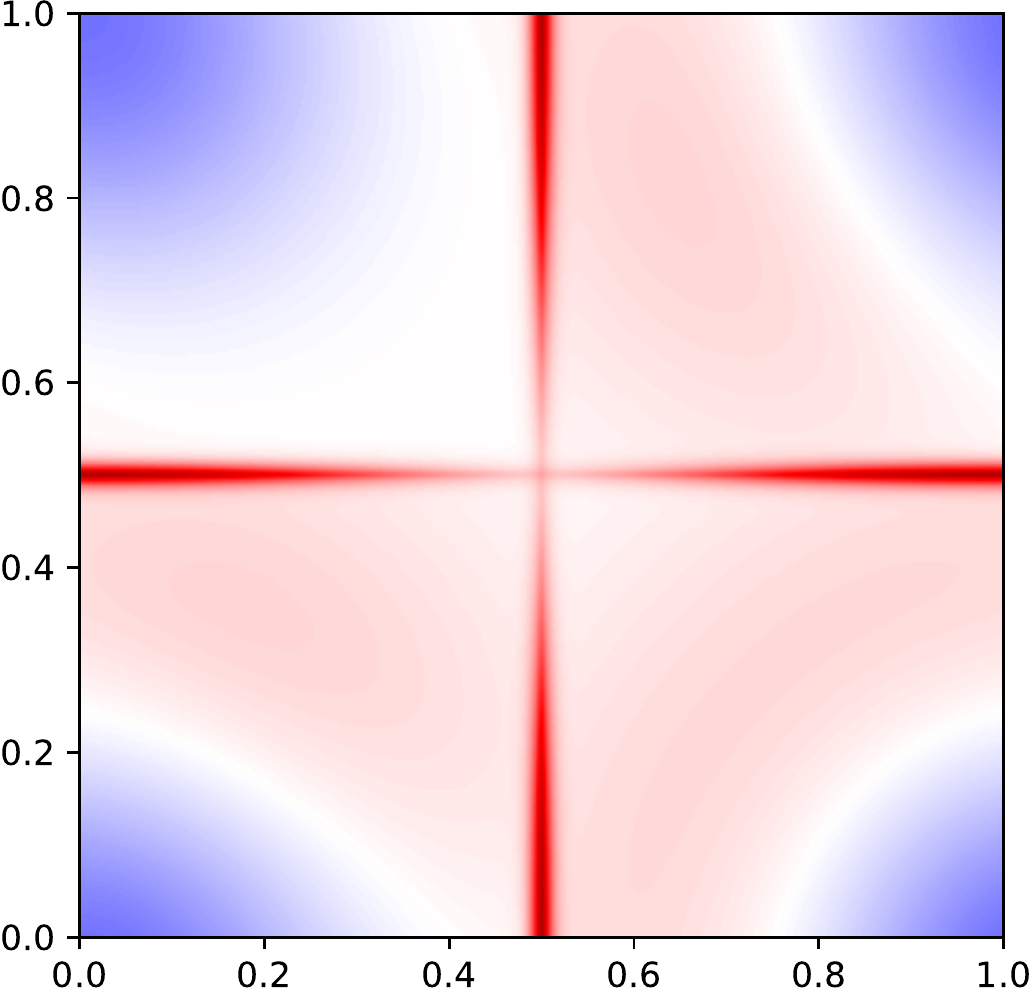}
      };
    \end{tikzpicture}
    \vspace*{13em}
    \caption{Full grid solver error $\ufgmax - \uana$}
    \label{subfig:advection:fg_error}
  \end{subfigure}%
  \hfill$+$\hfill
  \begin{subfigure}[]{0.2\textwidth}
    \centering
    \begin{subfigure}[]{\textwidth}
      \centering
      \begin{tikzpicture}[remember picture]
        \node (n1) at (0,0) {
          \includegraphics[width=\linewidth]{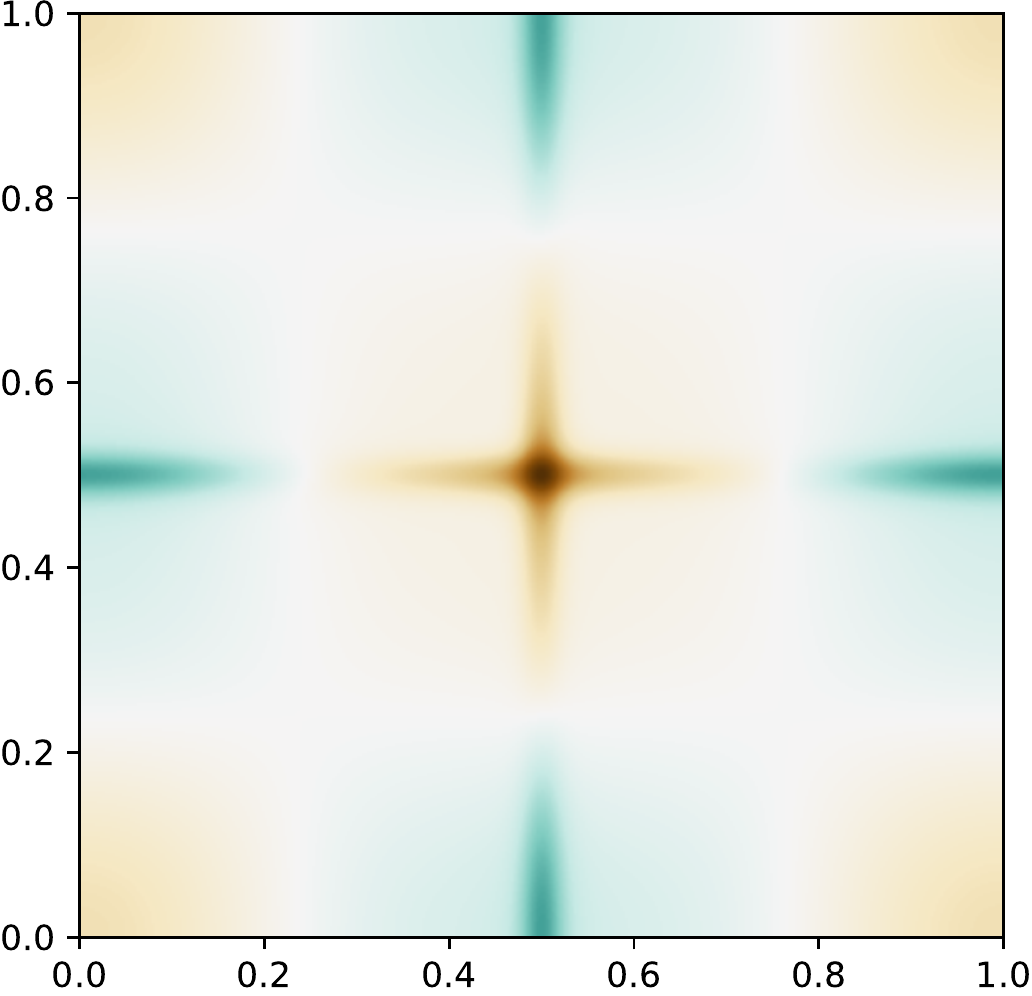}
        };
      \end{tikzpicture}
      \caption*{hierarchical hat basis \hbasishat}
      \label{subfig:advection:fg_hat_error}
    \vspace*{1em}
    \end{subfigure}
    \begin{subfigure}[]{\textwidth}
      \centering
      \begin{tikzpicture}[remember picture]
        \node (n1) at (0,0) {
          \includegraphics[width=\linewidth]{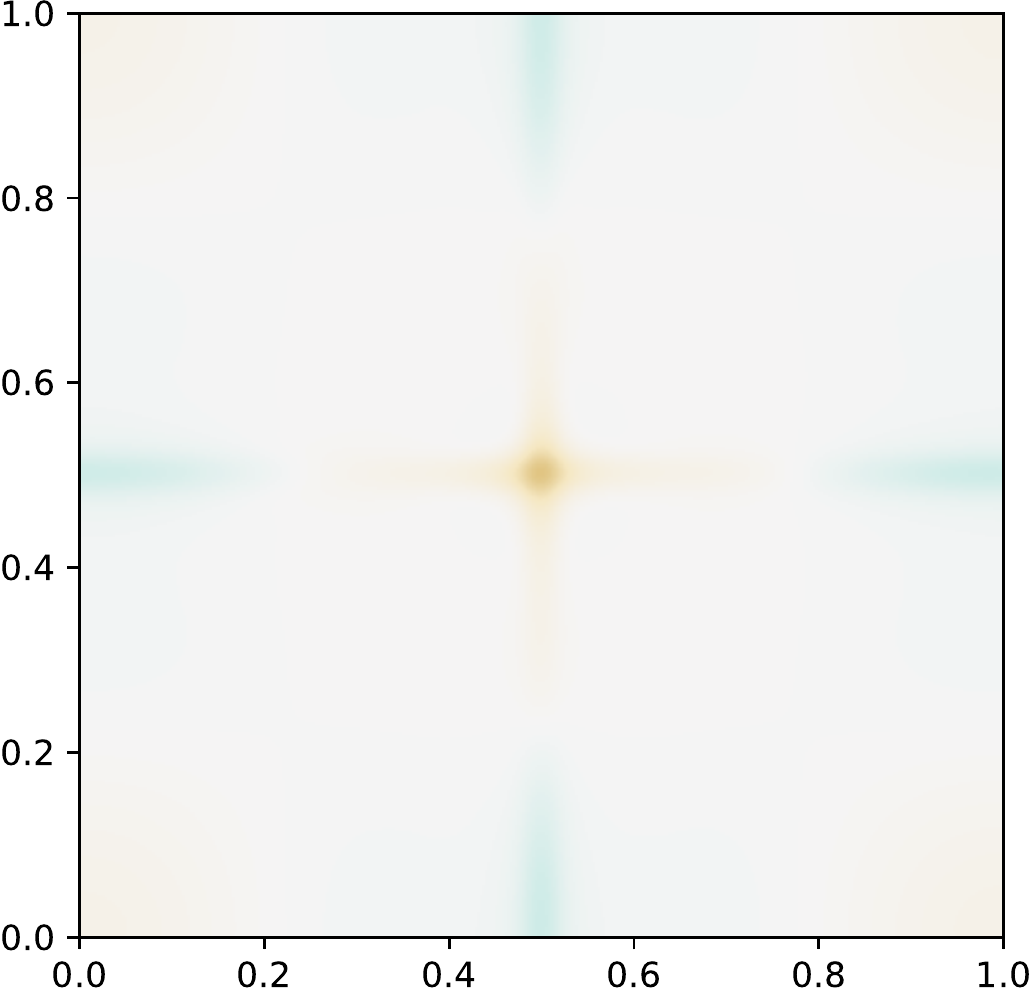}
        };
      \end{tikzpicture}
      \caption*{biorthogonal basis \hbasisbiorthogonal}
      \label{subfig:advection:fg_biorthogonal_error}
    \vspace*{1em}
    \end{subfigure}
    \begin{subfigure}[]{\textwidth}
      \centering
      \begin{tikzpicture}[remember picture]
        \node (n1) at (0,0) {
          \includegraphics[width=\linewidth]{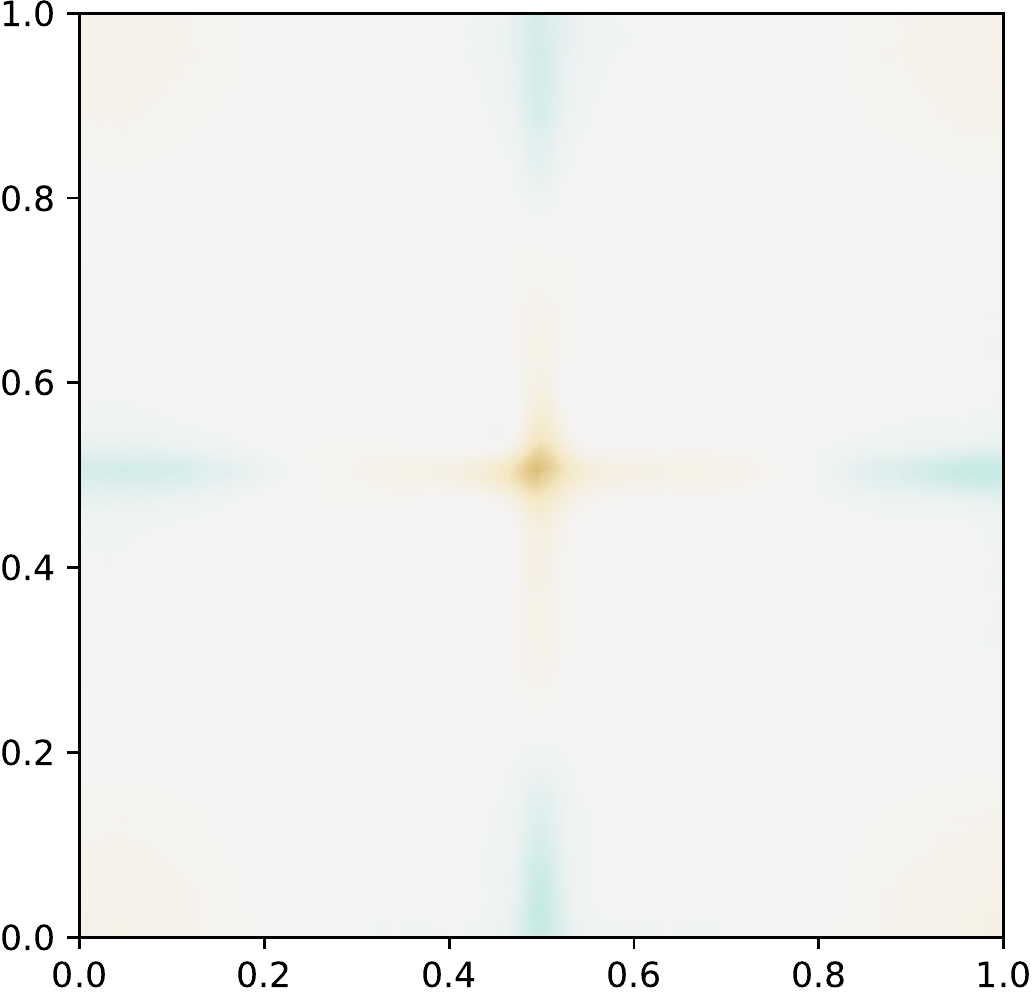}
        };
      \end{tikzpicture}
      \caption*{full weighting basis \hbasisfullweighting}
      \label{subfig:advection:fg_fullweighting_error}
    \end{subfigure}
    \caption{Error between full grid and CT solution ${\uct - \ufgmax}$}
    \label{subfig:advection:fg_ct_error}
  \end{subfigure}\hfill
  \begin{subfigure}[]{0.08\textwidth}
    \includegraphics[width=\textwidth]{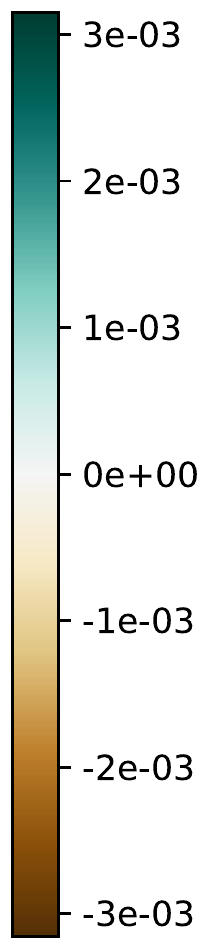}
  \end{subfigure}
  \caption{Error decomposition for the 2D advection solution at $t=0.5$: The combination technique error is split into the full grid error and the difference between full grid and combination technique solution.
  All solvers used $\lmax=(11, \,11)$, and the combination technique used $\lmin=(2, \,2)$, leading to approximately \SI[round-mode=figures,round-precision=3]{4.194304e6}{\dof} for the full grid and \SI[round-mode=figures,round-precision=3]{1.18784e5}{\dof} for the combination technique simulations, cf. the ellipses in \cref{fig:advection:errors}.
  We use $t=0.5$ instead of $t=1.0$, because the slight differences in \cref{subfig:advection:ct_error} are easier to compare when the maxima of the errors are centered.
  The difference between the full grid and CT solutions is visible on a smaller scale only, hence the separate color bar on the right.}
  \label{fig:advection:difference}
\end{figure}

These errors can be split up in a meaningful way:
For a combination scheme with minimum level $\lmin$ and maximum level $\lmax$, it seems unlikely to attain errors below that of the reference full grid solution \ufgmax, since all coordinates contained in the sparse grid are a subset of the full grid's \si{\dof}.
We therefore take the full grid errors as a bound to the combination technique errors.
Thus, we may decompose the total error into the solver's error itself and the error introduced by the combination technique scheme:
\begin{equation}\label{eqn:advection:errordecomp}
  \begin{split}
    \uct - \uana &= (\ufgmax - \uana) + (\uct - \ufgmax)\ , \\
    \norm{\uct - \uana} &\leq \norm{\ufgmax - \uana} + \norm{\uct - \ufgmax} \ .
  \end{split}
\end{equation}

Visually, it is instructive to investigate one fixed set of data points, namely the two-dimensional problem with maximum level $\lmax=(11,\,11)$ (\cf the ellipse in \cref{fig:advection:errors}).
The analytical solution at $t=0.5$, interpolated onto $\justl=\lmax$ (\ie $N = (2048,\,2048)$), is displayed in \cref{fig:advection:analytical}.

In \cref{fig:advection:analytical,fig:advection:difference}, the solutions are displayed at time $t=0.5$, \ie the Gaussian in the analytical solution has moved along the diagonal, such that the maximum of $1$ is now exactly on the corner points.
The error field w.r.t. the analytical solution is shown in \cref{fig:advection:difference}---making it a visualization of the error decomposition, \cref{eqn:advection:errordecomp}.
To plot the two-dimensional fields, we choose $t=0.5$ instead of $t=1.0$ (like in \cref{fig:advection:errors}) because the slight differences in \cref{subfig:advection:ct_error} are easier to compare when the maxima of the errors are centered.

For all solver schemes, most of the error is introduced along the middle axes.
This is to be expected, given that the analytical solution is $C^\infty$ in most of the domain, but only $C^0$ along the middle axes, since the middle axes at $t=0.5$ are the boundary of the initial solution $u(t=0)$ that is being advected through the domain.
High values of \uana are underestimated while low values tend to be overestimated.
We observe that the errors are dominated by the full grid error $\norm{\ufgmax - \uana}$, and the sparse grid projection error $\norm{\uct - \ufgmax}$ is more than one order of magnitude lower for all choices of hierarchical basis functions---note the different color scale in \cref{fig:advection:difference}.
Also, the sign of the sparse grid projection error appears to form a checkerboard pattern:
The values are underestimated at the center of the domain, partially compensating the full grid solver's error.
In the other regions of $\Omega$, the full grid solver's error is slightly amplified by the sparse grid projection error.

Yet, strikingly, the difference between the mass-conserving CT solutions and the full grid solutions $\norm{\uct - \ufgmax}$---the combination technique projection error---is significantly lower than for the hat basis combination technique: It generally has the same shape but is lower by a factor of $\approx 3$ throughout the domain.

This also explains the differences in the error norms highlighted in \cref{fig:advection:errors}:
For the same \lmax, the projection error measure $\norm{\uct - \ufgmax}_p$ is decreased by a factor of $\approx 3$ when using $\hbasisbiorthogonal$ or $\hbasisfullweighting$ instead of $\hbasishat$ for the combination.
This error reduction is observed in all advection simulations considered here---across all dimensionalities $d$ and norm parameters $p$.

In addition to maintaining the conservation of mass, the increased accuracy is a substantial benefit for combination technique simulations.

\subsection{Landau damping with the semi-Langrangian solver}\label{subsec:experiments:landau}

For the Vlasov--Poisson system, we study two typical benchmark problems, Landau damping and the two-stream instability.
Landau damping refers to the effect of exponential damping of longitudinal waves in plasmas.
It can be studied when simulating the Vlasov--Poisson system with the initial condition
\begin{equation}
f_{0}(\vect{x},\vect{v}) = \left( 1 + \varepsilon \sum_{i=1}^3 \cos(k x_i)\right) \frac{1}{(2\pi)^{3/2}} \exp\left( - \frac{\|\vect{v}\|_2^2}{2} \right)
\end{equation}
on the domain $[0, \frac{2\pi}{k}]^3 \times \R^3$ for certain ranges of $k$, typically $k=0.5$.

A linear dispersion analysis reveals the damping rate and the oscillation frequency in time to first order in $\varepsilon$.
For the mode $k=0.5$ the damping rate is $\omega = -0.1533$.
For a small value of $\varepsilon=0.01$, the electric energy practically coincides with the linear prediction from the second oscillation onwards.
In grid-based simulations, however, the so-called numerical recurrence phenomenon occurs: After a certain time depending on the resolution of the velocity grid, more precisely when $T = \frac{2\pi}{k \Delta v}$ with $\Delta v$ being the (minimal) resolution in velocity space, a sudden growth in potential energy occurs.
This phenomenon is a purely numerical artefact.

For low-resolution test runs of the combination technique, we could verify that the combined potential energy $\norm{\phi}^2_{2,\text{ct}}$ is a very good approximation to the sparse grid potential energy $\norm{\phi}^2_{2,\text{sg}}$, \cf \cref{eq:linearization}:
The sparse grid's energy $\norm{\phi}^2_{2,\text{sg}}$ is obtained by interpolating $f_{\text{ct}}$ onto \lmax and then computing $\norm{\phi}^2_{2}$ on the finely-resolved grid.
This holds at least for the regions where the simulation is stable---in those cases where $\norm{\phi}^2_{2,\text{ct}}$ becomes smaller than $0$, it is plausibly quite different from $\norm{\phi}^2_{2,\text{sg}}$, since the latter is guaranteed to be positive.

\begin{figure}[h!]
  \centering
\includegraphics[width=\linewidth]{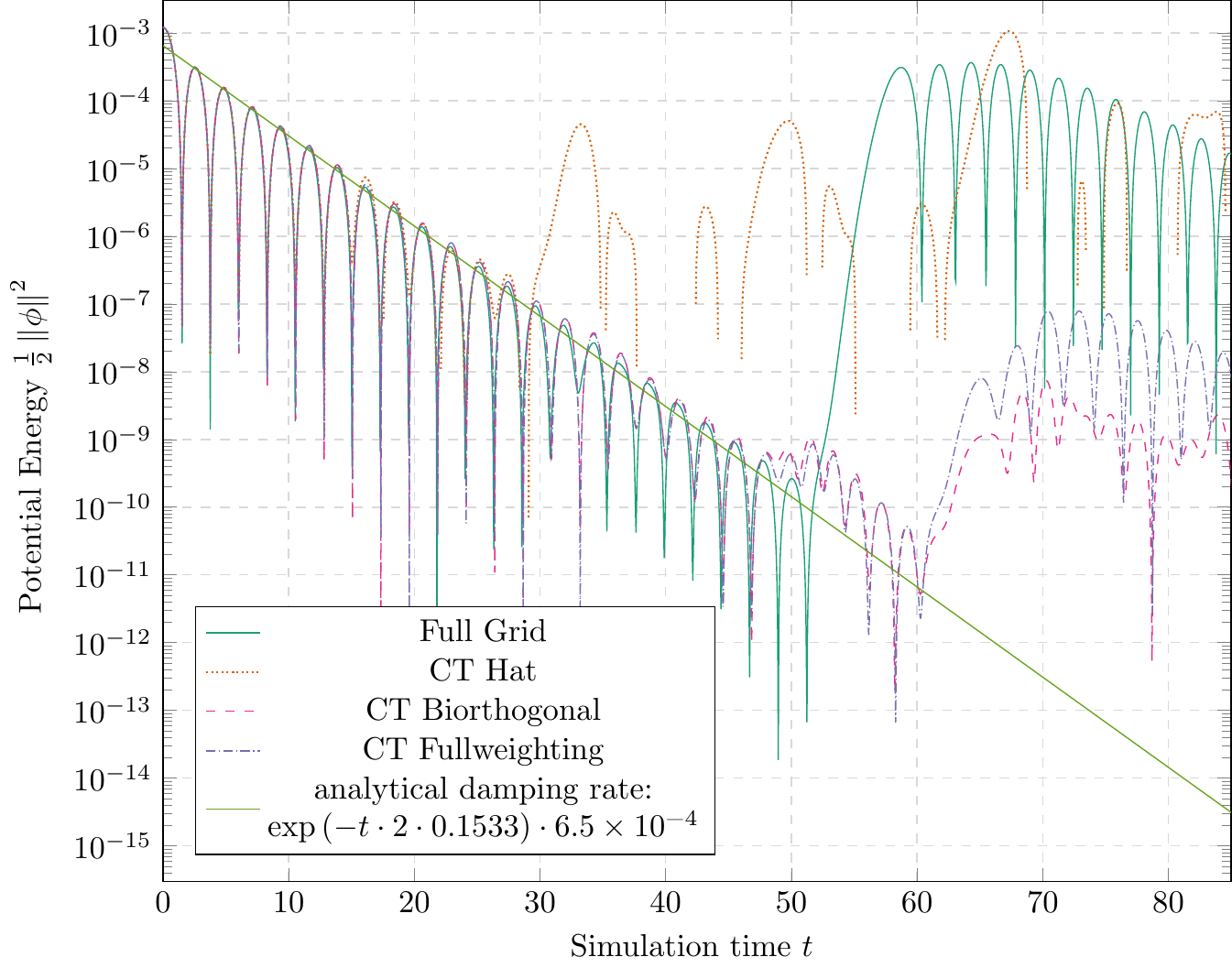}

  \caption{Landau Damping: The \SI{5.5}{\gibi\byte} full grid with a resolution of $\vect{N} = (16, \, 16, \, 16, \, 57, \, 57, \, 57)$ is compared to combination technique solutions with $\lmin = (4, \,4, \,4, \,2, \,2, \,2)$, $\lmax = (4, \,4, \,4, \,8, \,8, \,8)$ at \SI{5.4}{\gibi\byte} for the $f$ data structures.
  This means that the $\vect{x}$ space resolutions are exactly the same between the full and combination simulations, and combination is only employed in the $\vect{v}$ space, resulting in 64 component grids for the combination technique simulations.
  For the combination technique simulations, combination is performed after every time step of $\Delta t = 0.01$.
  }
  \label{fig:landau:B0:5GiB}
\end{figure}

Initial Landau damping test simulations showed that the hierarchical coefficients rapidly approach zero for finer $\vect{x}$ space resolutions, such that the $\vect{x}$ levels are uniformly chosen at a relatively coarse resolution of $\scalarl_\vect{x} = 4$ for this experiment.
This also becomes clear from the structure of the problem, since the perturbation is chosen aligned with the grid.
This alignement persists in the dynamics of the linearized equation that decouples the three one-dimensional perturbations.
Thus, we are effectively reducing the combination dimensions to the three velocity components $v_1, v_2, v_3$.

The exact combination technique scenario is $\lmin = (4, \,4, \,4, \,2, \,2, \,2)$, $\lmax = (4, \,4, \,4, \, 8, \,8, \,8)$, resulting in 64 component grids, which take up a total of \SI{5.4}{\gibi\byte} for the distribution functions' data structures.
We compare this to the same simulation run on a single full grid with resolution $\vect{N} = (16, \, 16, \, 16, \,  57, \, 57, \, 57)$ (i.e, $\justl \approx (4, \,4, \,4, \, 5.83, \,5.83, \,5.83)$), totalling $\approx \SI{5.5}{\gibi\byte}$ for the $f$ field and matching the $\vect{x}$ resolutions of the combination technique simulations exactly.

The simulation is performed with a three-point Lagrange interpolation used for the semi-Lagrangian method.
Using this relatively low-order interpolation, we have a minimal overhead due to ghost-cell transfer but on the other hand, we clearly see some deviations of the simulation from the analytical damping rate.
For the full grid solution, we expect the recurrence at time $T=\frac{2\pi \cdot 57}{0.5 \cdot 12} \approx 59.7$ which can also be observed (cf.~Fig.~\ref{fig:landau:B0:5GiB}).
The combination technique solutions show no clear recurrence but we observe smaller jumps in the damping profile of the energy curve.
With the pure hat function sparse grids, this occurs already very early around time 30.
For the full weighting and biorthogonal solutions, on the other hand, the effect occurs at a time similar to the recurrence time.

\subsection{Plasma instability with the semi-Langrangian solver}\label{subsec:experiments:instability}

Next, we consider the two-stream instability with initial condtion
\begin{equation}
\begin{aligned}
	f_{0}(\vect{x},\vect{v}) =& \left( 1+  \varepsilon \sum_{i=1}^3 \cos(0.2 x_i) \right)  \frac{1}{2(2 \pi)^{3/2}} \cdot \\
    &\left( \exp\left( - \frac{(v_1-2.4)^2}{2} \right) +  \exp\left( - \frac{(v_1+2.4)^2}{2} \right) \right) \exp\left( - \frac{v_2^2+v_3^2}{2} \right)
\end{aligned}
\end{equation}
on the domain $\left[0,\frac{2 \pi}{0.2}\right]^3 \times \R^3$.
A linear dispersion analysis shows that there is a growing mode with growth rate given by $\omega = 0.2258$.
The two-stream instability is bound to appear along the first dimension while Landau damping dominates along the others.
For a perturbation with $\varepsilon=0.001$, the potential energy curve follows the linear dispersion after some initial oscillations with a growth rate of $\omega$.
However, at some point the perturbation from equilibrium becomes so strong that nonlinear effects start dominating: Particles get trapped and the electric energy oscillates around a certain energy level.

Typical pitfalls of numerical methods during the nonlinear phase are too high oscillations or numerical damping of the electric energy, as observed in previous experiments~\cite{kormann_sparse_2016}.

\begin{figure}[p!]
  \centering
  \includegraphics[width=\linewidth]{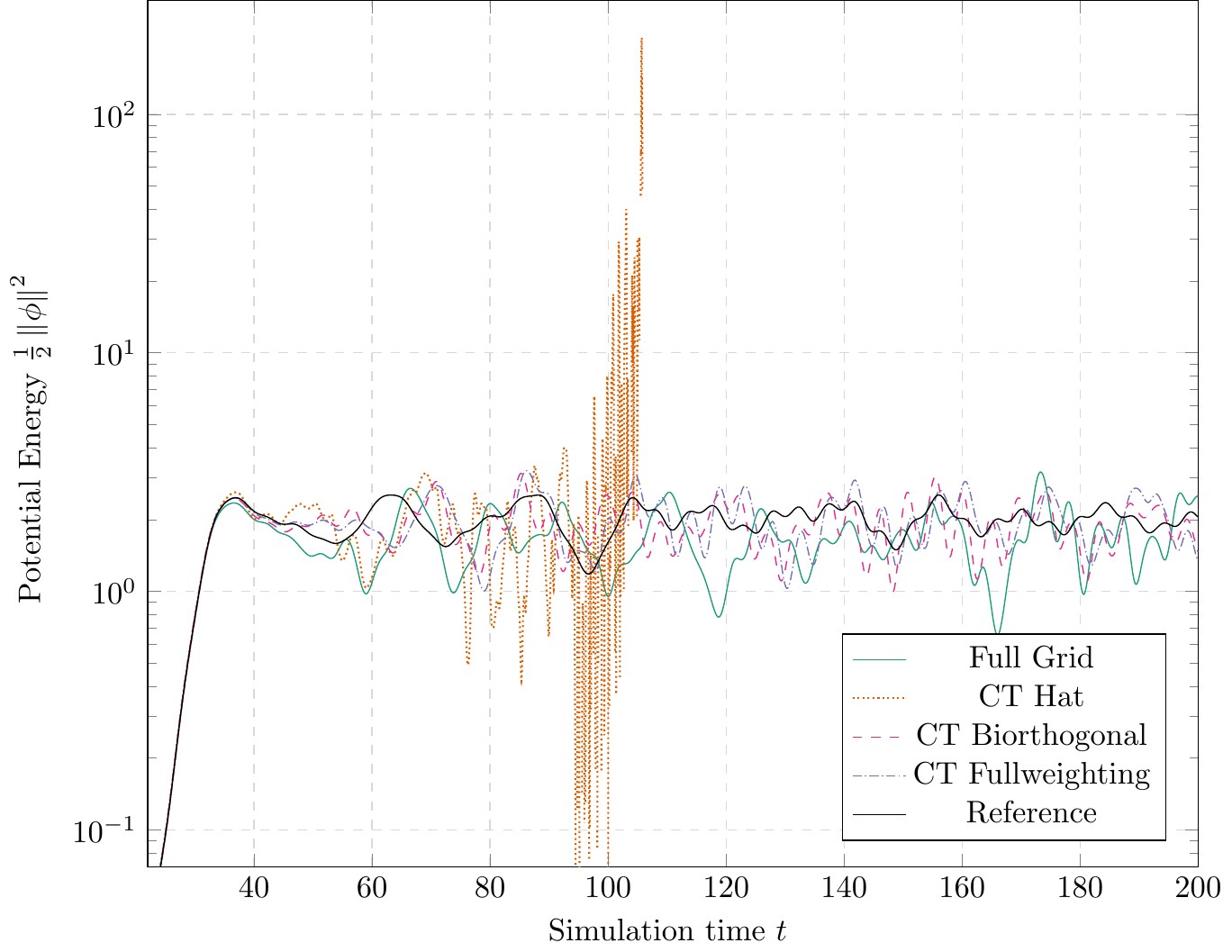}
 \caption{The potential energy over time in the two-stream instability scenario.
 The combination technique simulations use $\lmin=(3, \,3, \,3, \,3, \,3, \,3)$, $\lmax=(6, \,6, \,6, \,6, \,6, \,6)$ (84 component grids) and the full grid solution uses $\vect{N}=(22, \,22, \,22, \,24, \,24, \,24)$---both resulting in a memory footprint of $\approx\SI{1.1}{\gibi\byte}$ for the distribution function.
 The reference solution has a resolution level of $\lmax$, taking up \SI{512}{\gibi\byte} in memory for the plain $f$ data.
 For the combination technique simulations, combination is performed after every time step of $\Delta t = 0.01$.
 }
 \label{fig:instability:multidim:1GiB}
  \includegraphics[width=\linewidth]{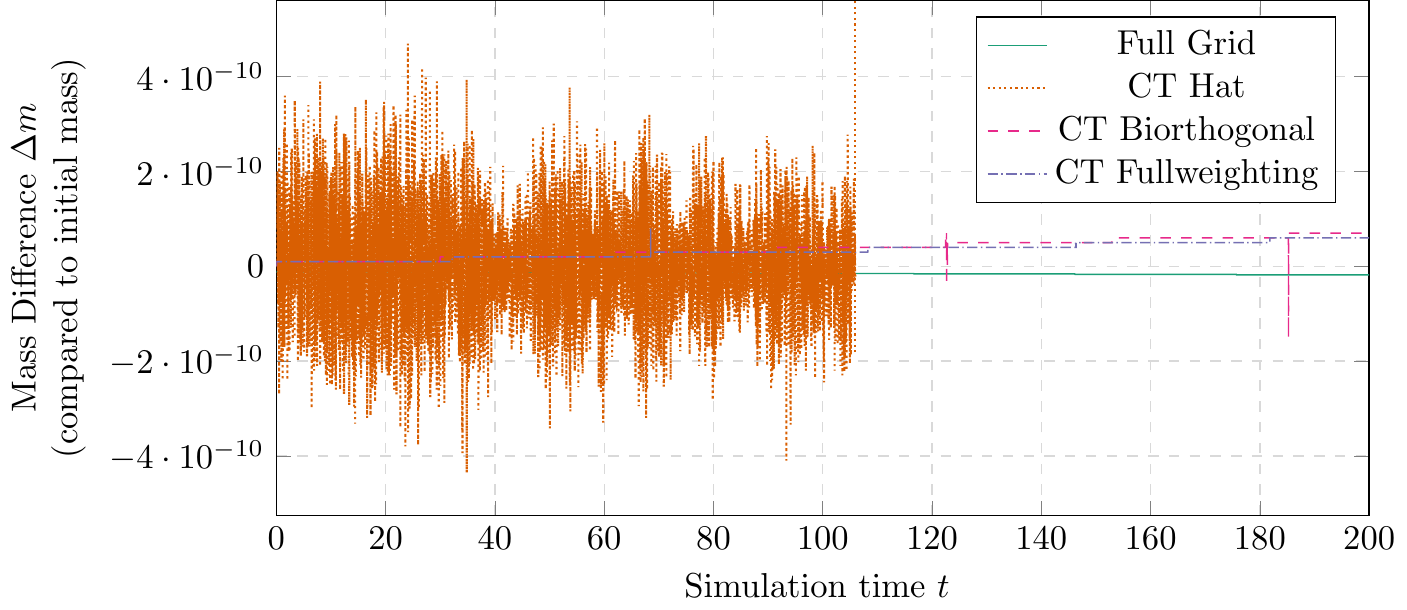}
  \caption{Difference between the mass $m_{\text{ct}}(t) = \sum_{\vect{l} \in \setI} \lambda_\vect{l} \cdot \int f_\vect{l}(x,t) \,dx$ and the initial mass $m_{\text{ct}}(0)$ for the two-stream instability scenario.}
  \label{fig:instability:multidim:1GiB:mass}
\end{figure}

Here, we compare a sparse grid combination solution with 84 component grids of $\lmin=(3, \,3, \,3, \,3, \,3, \,3)$, $\lmax=(6, \,6, \,6, \,6, \,6, \,6)$ and a single full grid solution with $\vect{N}=(22, \,22, \,22, \,24, \,24, \,24)$ (i.e, $\justl \approx (4.46, \,4.46, \,4.46, \, 4.58, \,4.58, \,4.58)$), since both configurations result in a similar memory footprint.
For the linear regime we recover the growth rate with very good accuracy in all simulations.
Therefore, in \cref{fig:instability:multidim:1GiB}, we focus on the nonlinear part of the simulation.
It can be seen that the pure combination technique solution based on hat functions becomes unstable just before time 106 while the other three simulations oscillate around the reference solution from a simulation on a full grid with $\vect{l} = \lmax$. In particular, no numerical damping is observed.
The medium-resolution full grid solution contains a rather uniform oscillation with a larger amplitude and can therefore be considered slightly worse than the ones obtained with a full weighting or a biorthogonal combination technique.

Figure \ref{fig:instability:multidim:1GiB:mass} shows the deviation of the mass over time.
It can be seen that the full grid solution conserves mass to machine precision and the full weighting and biorthogonal combination technique solutions show a slow and small variation in mass.
We attribute this to the accumulated errors from the individual hierarchization and dehierarchization operations on each grid:
By definition, the hierarchical increments $\alpha_{ls}$ should be very small in smooth regions of the solution.
In calculating them, we are subtracting very similar values from each other, which results in numerical cancellation for $\alpha$ to some degree.
The visible mass changes for the $\hbasisbiorthogonal$ and $\hbasisfullweighting$ at $\approx 10^{-11}$ are due to the output accuracy of 11 digits in the \selalib diagnostic files; the mass change slowly accumulates on each component grid but only becomes visible in \cref{fig:instability:multidim:1GiB:mass} once it has reached the output threshold of \num{1e-11} for the common sparse grid sum.
This is also how the tiny spikes are introduced at the changes of mass, for instance in the case of $\hbasisbiorthogonal$ at $t \approx 185$:
Some of the component grids display the change in mass one or two time steps earlier than the others, and it may happen that those grids' combination coefficients have the same sign, which leads to the spikes.
The combination coefficients also help to understand the high-frequent behavior when using $\hbasishat$ for the combination technique:
Even though the combination technique using the hierarchical hat basis is not mass-conserving, the error in mass shows an oscillatory behavior with rather small deviations up to the point where the simulation becomes unstable.
In particular, mass conservation remains on the order of $10^{-10}$ way beyond time $\approx 94$ where the values of the potential energy start to strongly oscillate, including negative values (which may occur due to the extrapolation of the potential energy from the values of the component grids, cf. \cref{eqn:advection:ctnorm}).

Thus, it appears that the conservation of mass itself is not the determining factor for the numerical stability of the problem.
Rather, the causality may be in the reverse direction: The choice of stable basis functions also introduces conservation of mass in the distribution function.

\section{Discussion of kinetic CT simulations with mass-conserving basis functions}\label{sec:discussion}

In summary, this work introduces the mass-conserving basis functions \hbasisbiorthogonal and \hbasisfullweighting for combination technique simulations.
While the biorthogonal basis has been previously used for sparse grid simulations~\cite{koster_multiskalen-basierte_2002}, the full weighting basis is a new approach that shows a connection to geometric multigrid methods.

A starting point for our research was the conservation of mass in the individual component grids, which is generally not to be expected if the combination technique is performed with hat functions.
The present work shows the conservation of mass by $\hbasisbiorthogonal$ and $\hbasisfullweighting$ theoretically as well as experimentally for a finite volume advection solver and the semi-Lagrangian \selalib solver, which both use periodic boundary conditions.
For solvers with non-periodic boundary conditions, theory tells that also the first moment---the location of the center of mass---should be conserved, which corresponds to conservation of momentum when considering velocity coordinates like in \selalib and other grid-based Vlasov solvers.

Furthermore, the direct comparison to the full grid solution in \cref{subsec:experiments:advection} showed that the mass-conserving basis functions decrease the projection error of the combination technique solution by a factor of $\approx 3$ as opposed to hat functions.

Perhaps the most important feature of the introduced mass-conserving bases is that they numerically stabilize the simulation.
In the Landau damping test case, \cf \cref{subsec:experiments:landau}, the standard combination technique with the hat basis loses numerical stability even before the first numerical recurrence occurs.
This problem is not encountered for the mass-conserving bases:
When allocating the same number of \si{\dof}, the biorthogonal and full weighting combination technique solutions are very similar to the full grid solution for the time before recurrence.

Fre the plasma instability scenario discussed in \cref{subsec:experiments:instability} the hat basis leads to the simulation becoming numerically unstable after some simulation time, which even causes it to abort.
By contrast, the biorthogonal and full weighting functions let the simulation run stable well into the quasi-stationary turbulent phase.
This fact is particularly interesting in the context of previous work in the EXAHD project~\cite{pfluger_exahd_2014,lago_exahd_2020,obersteiner_spatially_2021} which was severely impacted by recombination simulations with hat functions becoming numerically unstable.
Those cases that were most likely to become unstable were the ones that had many grids, fine resolutions, and short recombination intervals~\cite{obersteiner_spatially_2021}.
The results of the plasma scenarios in \cref{subsec:experiments:landau,subsec:experiments:instability} show that the mass-conserving basis functions solve this problem, at least for the range of resolutions that was tested.
It is worth noting that, contrary to previous numerical experiments directly in the hierarchical hat basis~\cite{kormann_sparse_2016}, no additional (unphysical) damping occurs in the two-stream instability scenario.
Further investigations are required to apply the approach to other grid-based solvers, in particular solvers that do not conserve any moments of the distribution function.

The time-based comparison of mass and potential energy in \cref{subsec:experiments:instability} suggests that the conservation of mass and the increased stability of the basis are closely linked: both result from the increased regularity of the biorthogonal wavelets $\tilde{\psi}$~\cite{cohen_biorthogonal_1992}.
Analytical results obtained by \citeauthor{koster_multiskalen-basierte_2002}~\cite{koster_multiskalen-basierte_2002} had predicted insufficient numerical stability for some PDE computations directly in the hierarchical hat basis sparse grid representation.
The same effect appears to hold in the \selalib simulations presented here, although the time-stepping does not take place in the hierarchical basis, but in the nodal representation on each component grid individually.
Only the numerical \enquote{synchronization} is carried out in the hierarchical basis, yet this is enough to skew the component solutions in a way that makes the simulation unstable over time.

Thus far, we observed no reason to prefer one of the two mass-conserving multiscale bases over the other.
The $2$D errors plotted in \cref{subsec:experiments:advection} illustrate that there is a slight difference in the shape of the error introduced, although it is very small and not distinguishable in the measured Monte Carlo error plots.
Nevertheless, the long-term behavior in \cref{fig:instability:multidim:1GiB} shows that these ever-so-slight discrepancies lead to differences in the time trajectory of the system.
At this point, it is not obvious whether one of $\hbasisbiorthogonal$ and $\hbasisfullweighting$ should be preferred over the other for combination technique simulations, but they both clearly outperform $\hbasishat$ in terms of conservation of mass, accuracy, and stability of the solution.

\section{Benefits and potential of mass-conserving basis functions for combination technique simulations}\label{sec:conclusion}

From our numerical experiments, it can be concluded that the newly introduced full weighting basis and the biorthogonal basis positively influence the numerical properties of the combination technique applied to high-dimensional advection problems compared to the state-of-the-art hat functions.

It is worth noting that the kinetic simulations performed in \cref{sec:experiments} use relatively simple, \enquote{standard} combination schemes.
By employing dimensionally~\cite{gerstner2003dimension,ruttgers_multiscale_2018} or spatially~\cite{obersteiner_generalized_2021} adaptive schemes, the accuracy could be improved further for the same number of degrees of freedom.
This applies to both selecting the initial set of component grid resolutions as well as automatically updating the scheme during the course of a simulation, in analogy to the adaptivity in the hierarchical bases as described in \cite{guo_adaptive_2017}.
Furthermore, the benefit of mass conservation is not limited to hyperrectangles. As long as the domain transform's Jacobian can be formulated in terms of a tensor product (such as for cylindrical coordinates), a straightforward transformation of the basis functions is possible.

And even for lower-dimensional simulations, it can be sensible to employ the combination technique with $\hbasisbiorthogonal$ and $\hbasisfullweighting$:
If the combination technique does not introduce any additional numerical instabilities, as is possible with the basis functions introduced in this work, it can be a viable option for finely resolved simulations on 2D or 3D structured grids.

The computational overhead introduced by the mass-conserving hierarchization and dehierarchization operations affects the additional parallelism provided by the combination technique only to a small extent~\cite{pollinger_scaling_2022}.
This fact allows to revisit previous attempts at exa-scale combination technique plasma scenarios~\cite{Obersteiner2017,heene_efficient_2016} with the newly introduced basis functions.
Even though codes such as \gene do not generally conserve mass, the increased accuracy and stability properties should be beneficial for the simulation of experimental nuclear fusion devices.
This includes results for algorithm-based fault tolerance in the presence of hard and soft faults~\cite{Obersteiner2017} as well as the distribution across compute systems~\cite{pollinger_distributing_2021}.

Furthermore, the concept of biorthogonal basis functions shows a way of extending the conservation to higher-order moments of the combined function, such as velocity moments of the Vlasov distribution function.
To this end, it is necessary to use higher-order ansatz functions in the simulations, such as a B-spline basis~\cite{Valentin2019}.
Then, longer filters $h,\tilde{h},g,\tilde{g}$ can be applied to retain more moments~\cite{cohen_biorthogonal_1992} of the combined function.

All of these properties---stability, conservation, adaptivity, fault tolerance, and parallelizability---are highly desirable for predictive plasma turbulence simulations at scale.
For grid-based solvers, the combination technique offers a black-box approach to break the curse of dimensionality, without the complexity of involved multiscale sparse grid solver schemes.
In this way, the mass-conserving hierarchical basis functions will hopefully foster the usage of the combination technique in plasma (and other) simulations.

\section*{Reproducibility references}

The simulations were generated using the codes \discotec (\url{https://github.com/SGpp/DisCoTec}) and \selalib(\url{https://github.com/selalib/selalib}).
The exact build instructions, parameters and outputs used for this work are made available at \url{https://doi.org/10.18419/darus-2790}.

\section*{Acknowledgments}
The authors would like to thank Michael Obersteiner for his help in coupling the codes \discotec and \selalib and for fruitful discussions. Discussions with Tilman Dannert in an early phase of the project are gratefully acknowledged.  Furthermore, the authors thank Michael Griebel for proofreading and helpful comments on the numerics of the combination technique.

\bibliography{refs}

\end{document}